\documentclass{aa}  
\usepackage{graphicx}
\usepackage{natbib}
\bibpunct{(}{)}{,}{a}{}{;}
\usepackage{amssymb}
\usepackage{txfonts}
\begin{document}
   \title{APEX-CHAMP$^+$ high-J CO observations of low-mass young stellar objects: I. The HH 46 envelope and outflow }


   \author{T.A. van Kempen
          \inst{1,2}
	  \and
	  E.F. van Dishoeck\inst{1,3}
	       \and
               R. G{\"u}sten\inst{4}
               \and
               L.E. Kristensen\inst{1}
               \and
           P. Schilke\inst{4}
       \and
        M.R. Hogerheijde\inst{1}
      \and 
      W. Boland\inst{1,5}
      \and
     B. Nefs\inst{1}
       \and
       K.M. Menten\inst{4}
      \and
      A.Baryshev\inst{6}
   \and
   F. Wyrowski\inst{4}}
   
          \offprints{T. A. van Kempen}

   \institute{$^1$ Leiden Observatory, Leiden University, P.O. Box 9513,
              2300 RA Leiden, The Netherlands\\	      
             $^2$ Center for Astrophysics, 60 Garden Street, Cambridge, MA 02138, USA      \\
 $^3$ Max-Planck Institut f\"ur Extraterrestrische 
             Physik (MPE), Giessenbachstr.\ 1, 85748 Garching, Germany \\
              $^4$ Max Planck Institut f\"ur Radioastronomie, Auf dem H\"ugel 69, D-53121, Bonn, Germany\\ 
             $^5$ Nederlandse Onderzoeksschool Voor Astronomie (NOVA), P.O. Box 9513, 2300 RA Leiden, The Netherlands \\
             $^6$ SRON Netherlands Institute for Space Research , P.O. Box 800, 9700 AV Groningen, The Netherlands\\
\email{tvankempen@cfa.harvard.edu}
                     }

   \date{Draft: 0.7 July 2008}
\titlerunning{The envelope, outflow and surroundings of HH~46}%

\def\placeFigureChapterFiveOne{
\begin{figure*}[!th]
\begin{center}

\includegraphics[width=140pt]{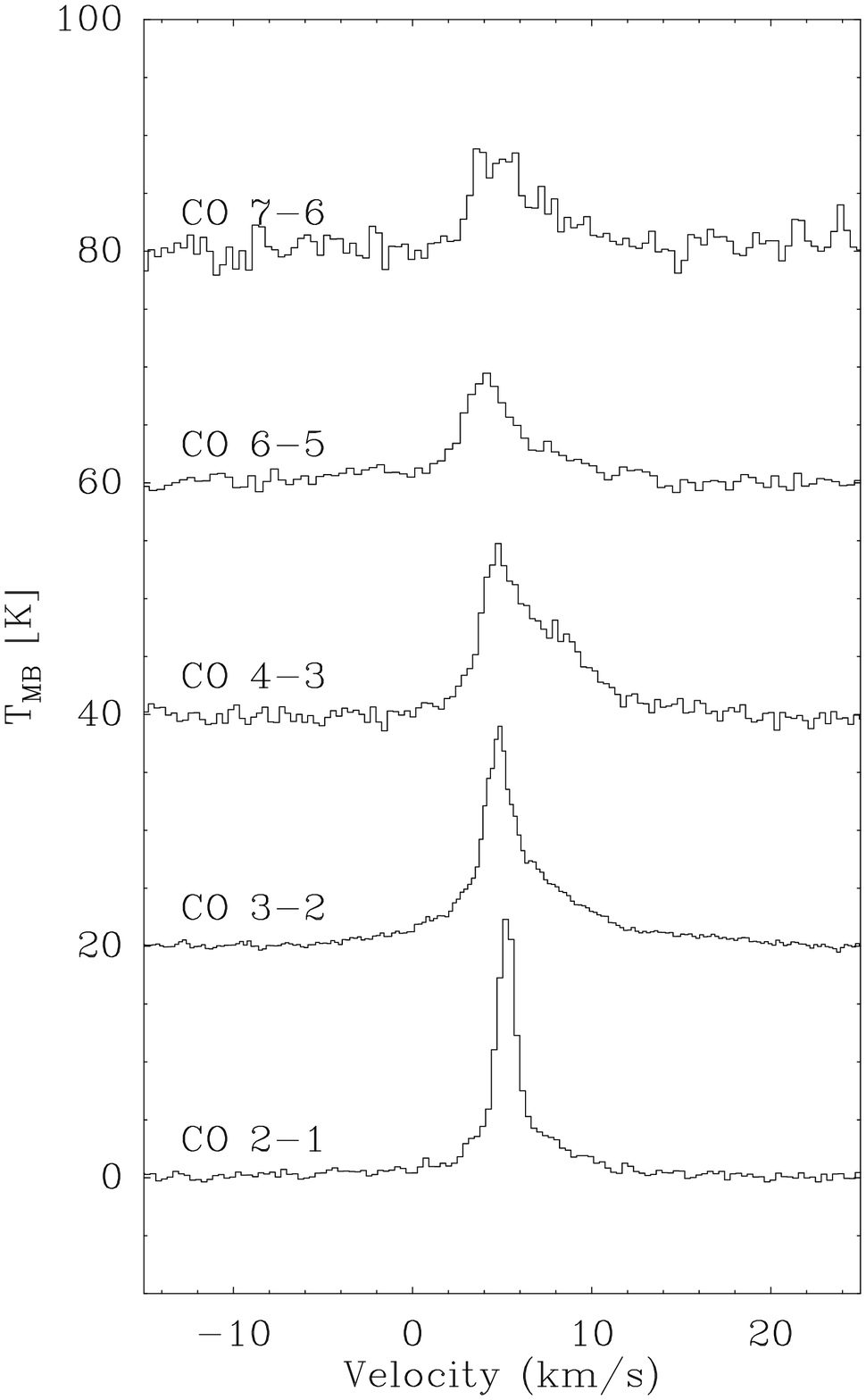}
\includegraphics[width=140pt]{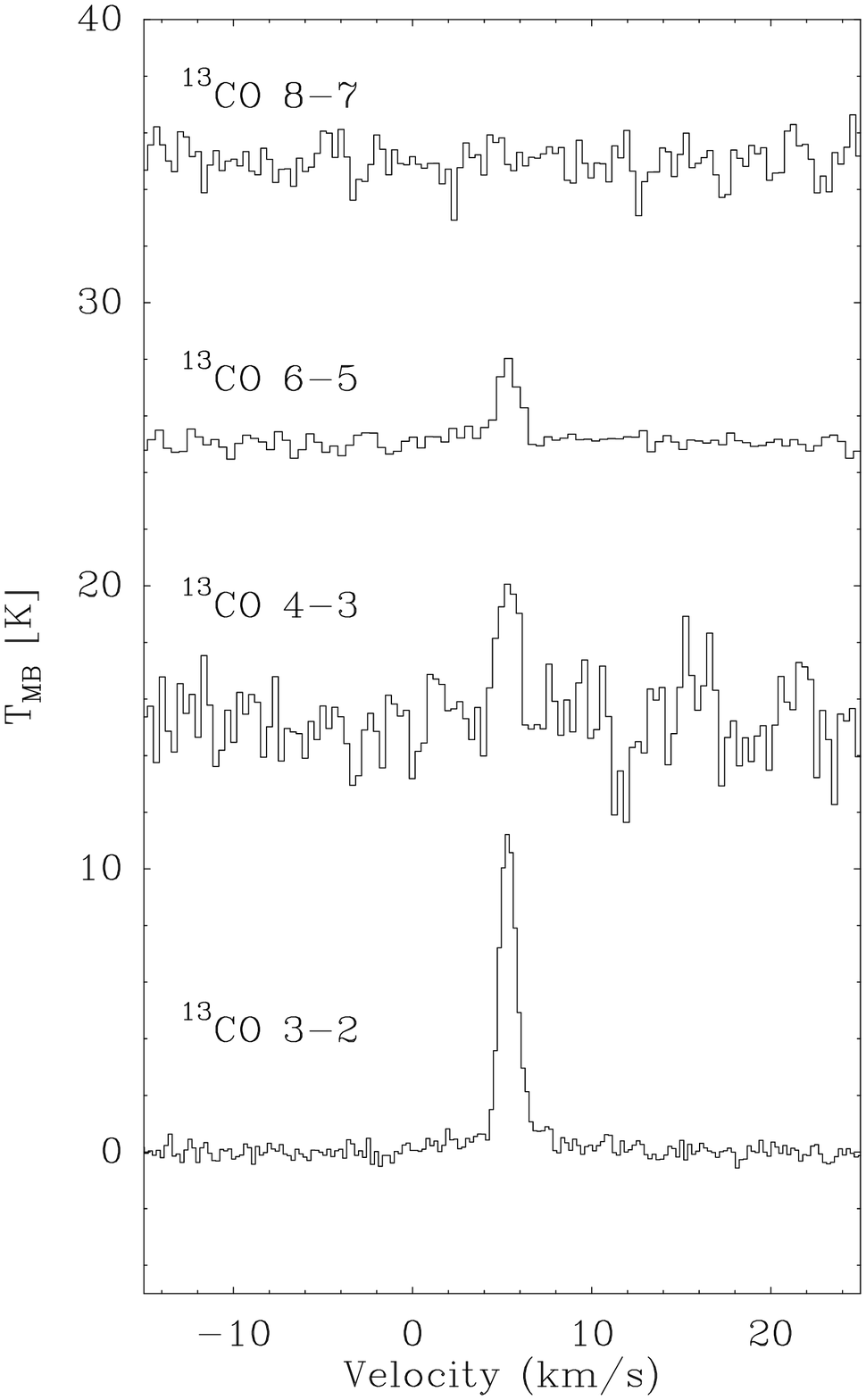}
\includegraphics[width=140pt]{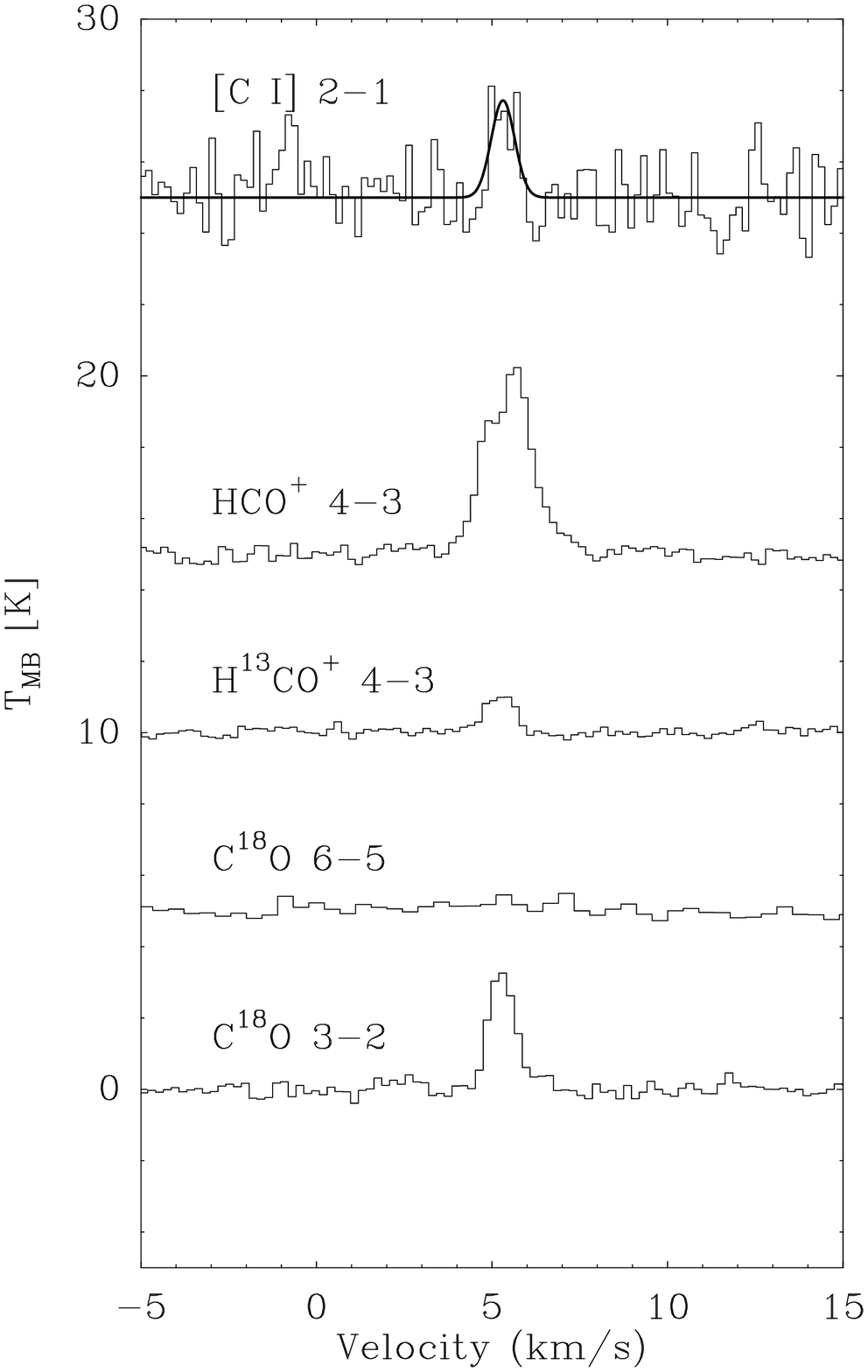}
\end{center}
\caption{Single spectra taken at the central position of HH~46 (all in order from {\it bottom} to {\it top}). {\it Left} : $^{12}$CO 2--1, $^{12}$CO 3--2, $^{12}$CO 4--3, $^{12}$CO 6--5 and $^{12}$CO 7--6. {\it Middle} : $^{13}$CO 3--2, $^{13}$CO 4--3, $^{13}$CO 6--5 and $^{13}$CO 8--7. {\it Right} : C$^{18}$O 3--2, C$^{18}$O 6--5, H$^{13}$CO$^+$ 4--3, HCO$^+$ 4--3 and [C I] 2--1. Spectra have been shifted vertically for easy viewing. }
\label{5:fig:spec_2a}
\end{figure*}
}

\def\placeFigureChapterFiveTwo{
\begin{figure}[!th]
\begin{center}
\end{center}
\includegraphics[width=240pt]{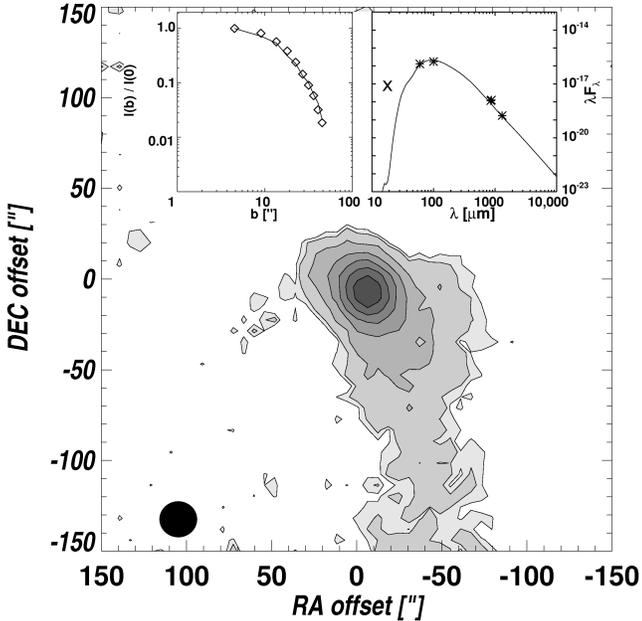}
\caption{Continuum map at 870 $\mu$m of HH~46 obtained with LABOCA
with contour levels plotted in multiples of the $3\sigma$ noise level
of 0.05 Jy/beam.  {\it Left inset}: the radial profile going outwards
from the IR position. {\it Right inset}: the SED from HH~46 from 10
$\mu$m to 10 mm. In both panels, the best-fitting DUSTY model is
overplotted. }

\label{5:fig:laboca}
\end{figure}
}

\def\placeFigureChapterFiveThree{
\begin{figure*}[!th]
\begin{center}
\includegraphics[width=140pt]{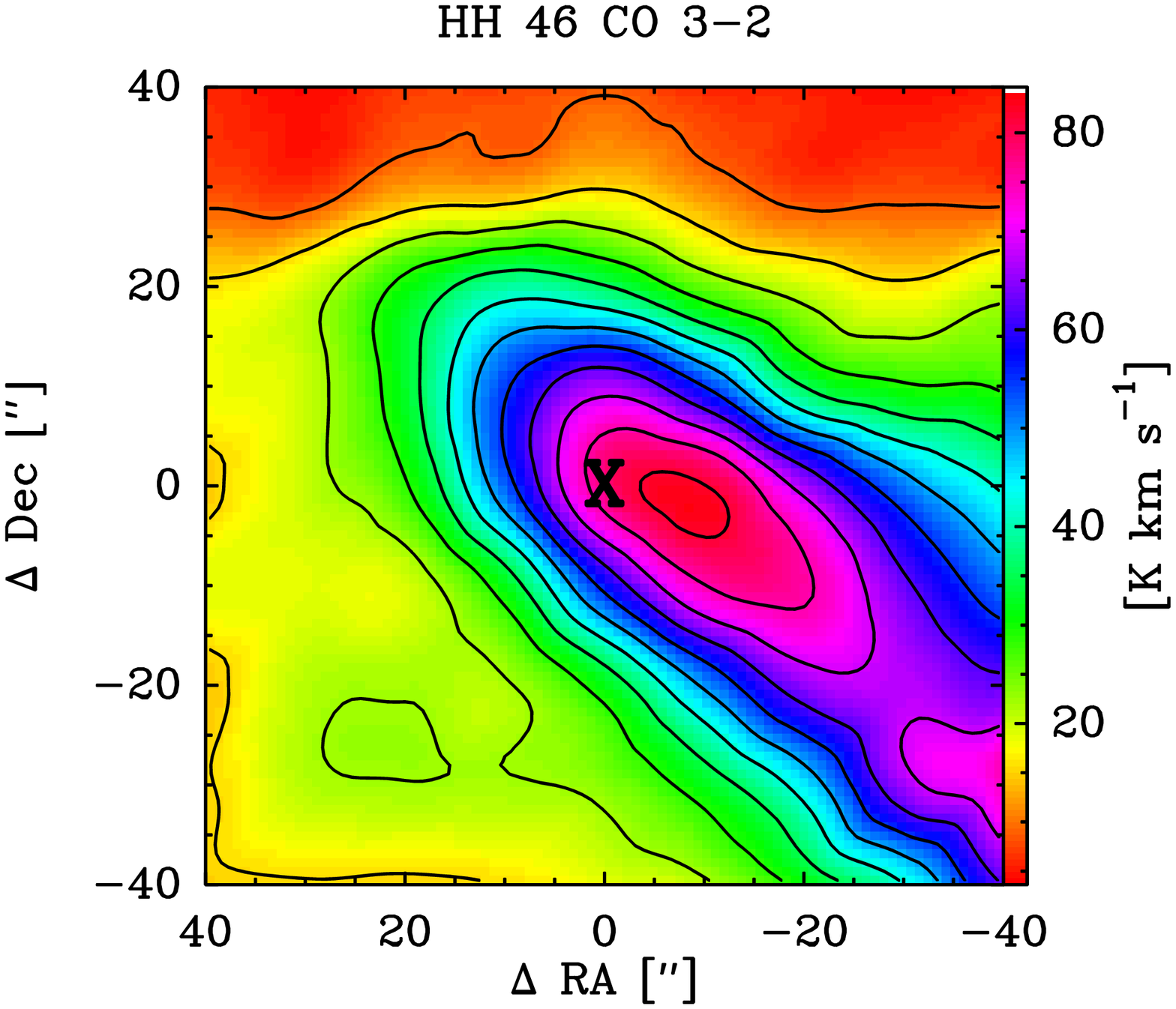}
\includegraphics[width=140pt]{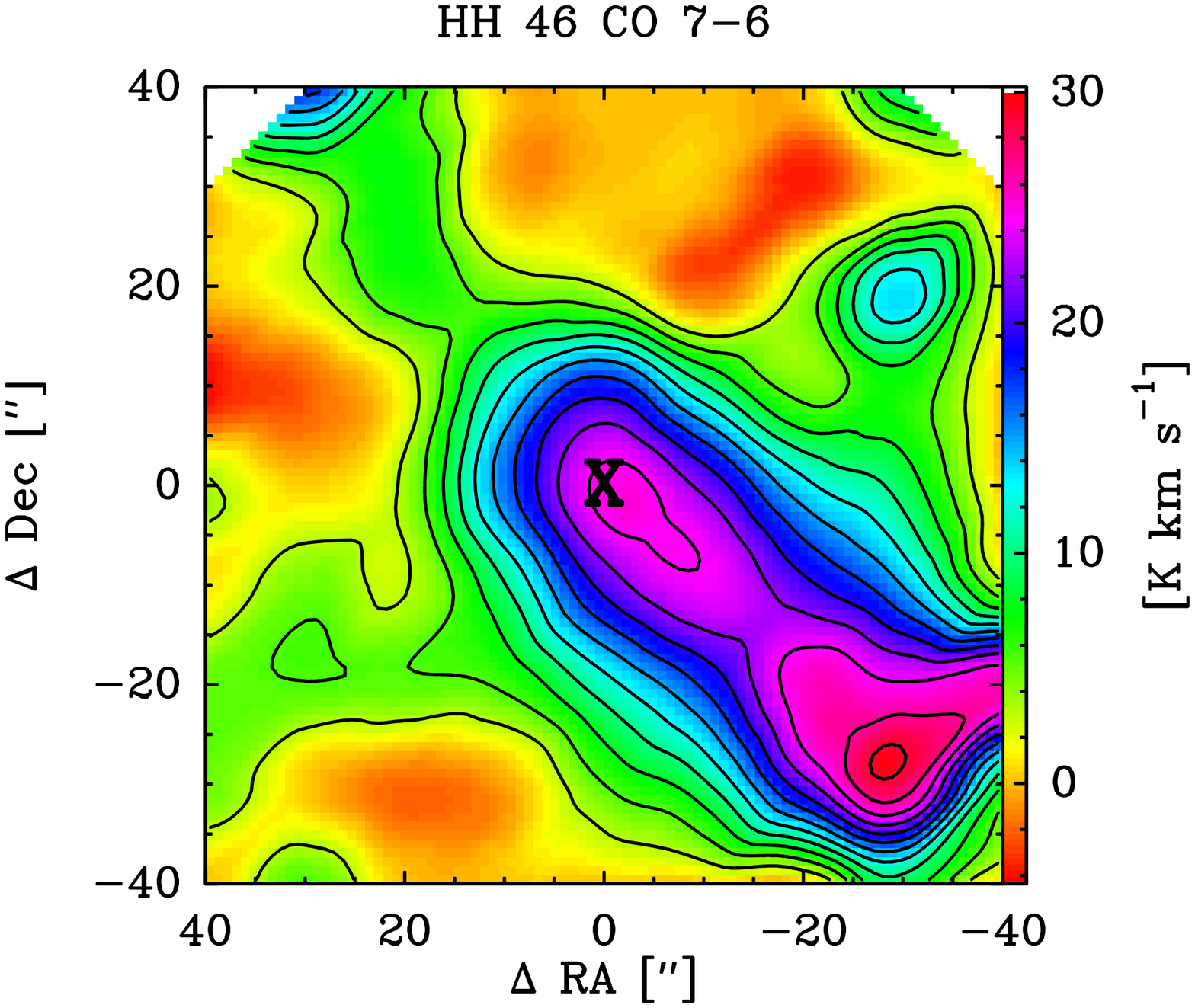}
\includegraphics[width=140pt]{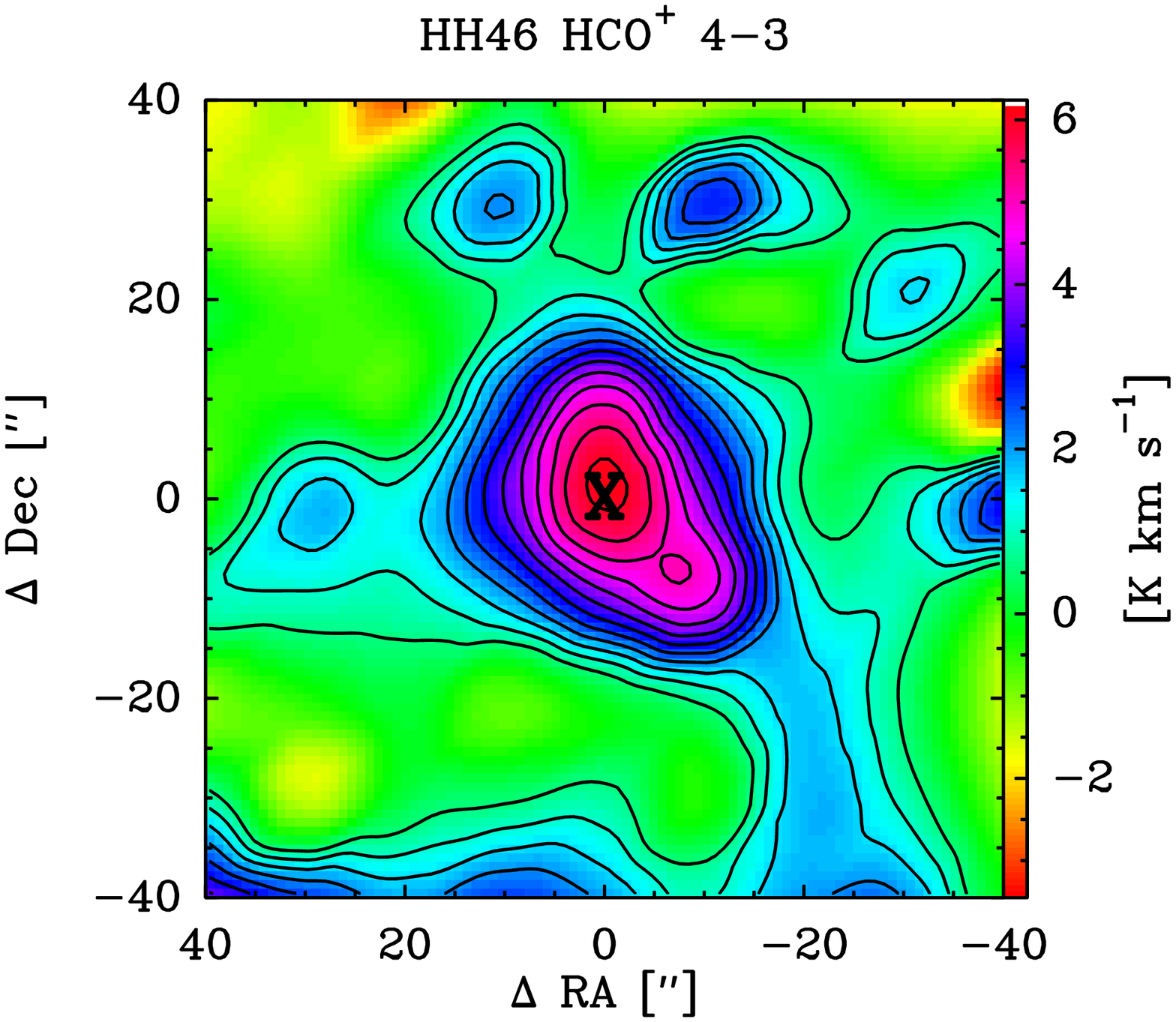}
\end{center}
\caption{From {\it left} to {\it right} : Spectrally integrated CO
3--2, 7--6 and HCO$^+$ 4--3 maps of HH~46 (see Table \ref{5:tab:line_res}). Contour levels are at
3$\sigma$, 6$\sigma$, 9$\sigma$, ... ($\sigma$= 1.5 K for CO 3--2, 0.4 K for HCO$^+$ 4--3 and 0.8 K for CO 7--6) The position of the IR source is
marked with a {\bf X}. Note that the HCO$^+$ 4--3 map has higher noise
levels than the central spectrum in Fig \ref{5:fig:spec_2a}. }
\label{5:fig:maps2a}
\end{figure*}
}

\def\placeFigureChapterFiveFour{
\begin{figure*}[!th]
\begin{center}
\includegraphics[width=450pt]{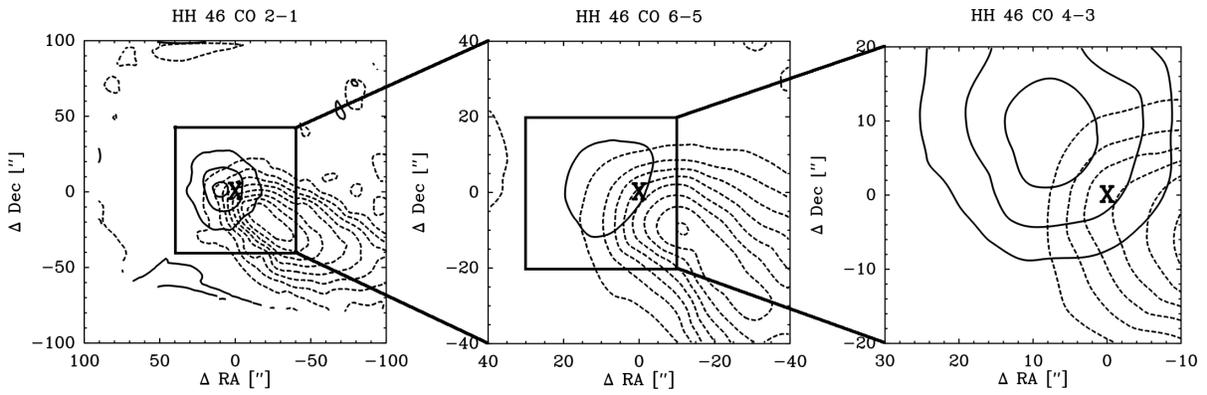}
\end{center}
\caption{Maps of HH~46 showing the outflowing gas. Solid lines show
blue-shifted, dashed lines how red-shifted emission. Contours are
drawn at 3$\sigma$, 6$\sigma$, 9$\sigma$,... , with the exception of
CO 2--1, where contours are drawn 3$\sigma$, 9$\sigma$, 15$\sigma$,
... . 3$\sigma$ levels are 0.4 K km s$^{-1}$, 1 K km s$^{-1}$ and 1 K
km s$^{-1}$ for 2--1, 6--5 and 4--3 respectively. Blue- and
red-shifted emission is derived by subtracting a central gaussian
fitted to the quiescent central part of the line profiles. Velocities
typically range from -5 to 1.5 km s$^{-1}$ for the blue-shifted
emission and 8 to 15 km s$^{-1}$ for the red-shifted. Note that the
images are at different spatial scales and progressively zoom in onto
the central protostar from left to right. The position of the infrared
source is marked in all maps with a {\bf $`$X$'$}.}
\label{5:fig:mapsvel}
\end{figure*}
}
\def\placeFigureChapterFiveFive{
\begin{figure*}[!th]
\begin{center}
\includegraphics[angle=270,width=180pt]{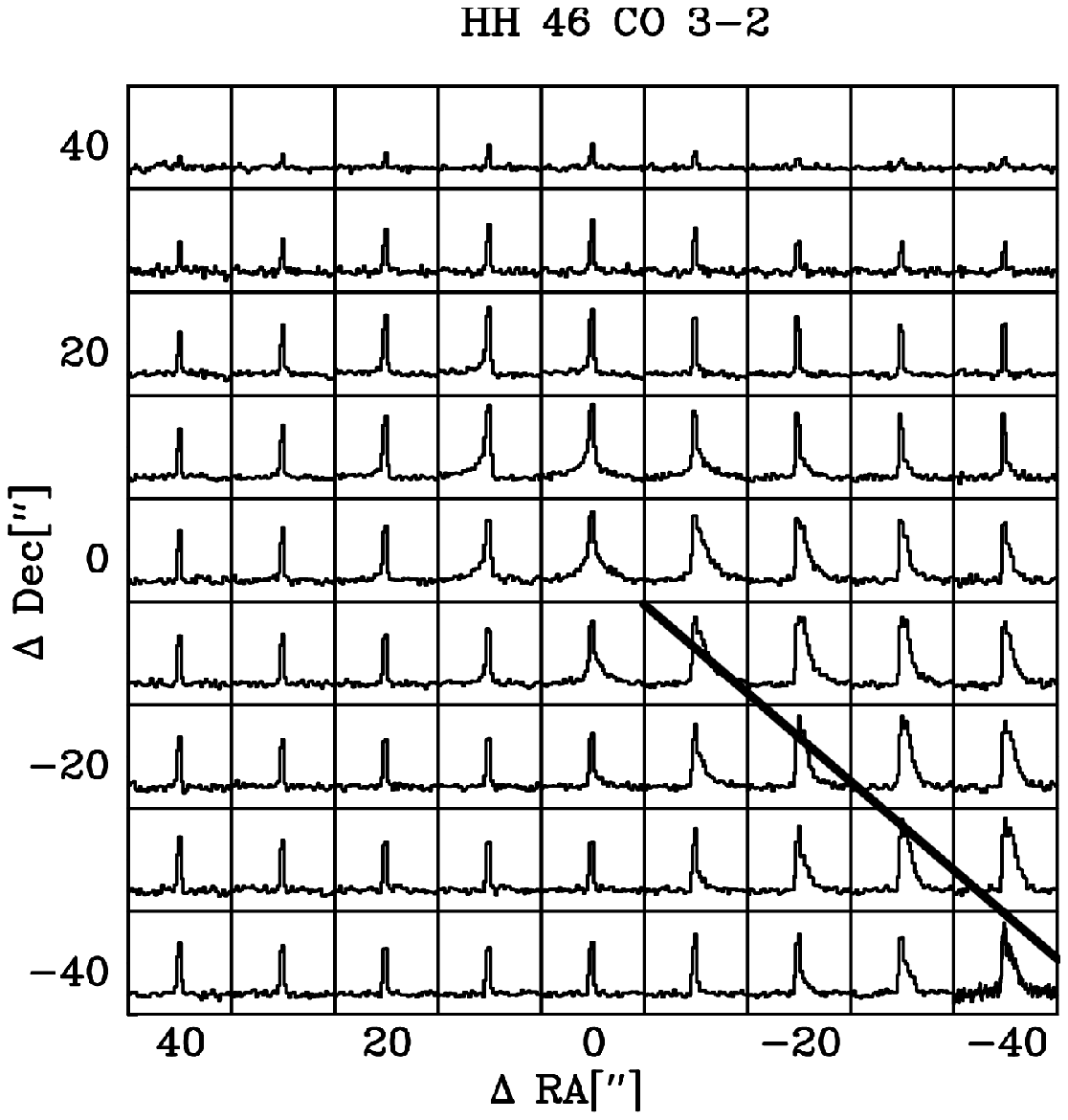}    
\includegraphics[angle=270,width=180pt]{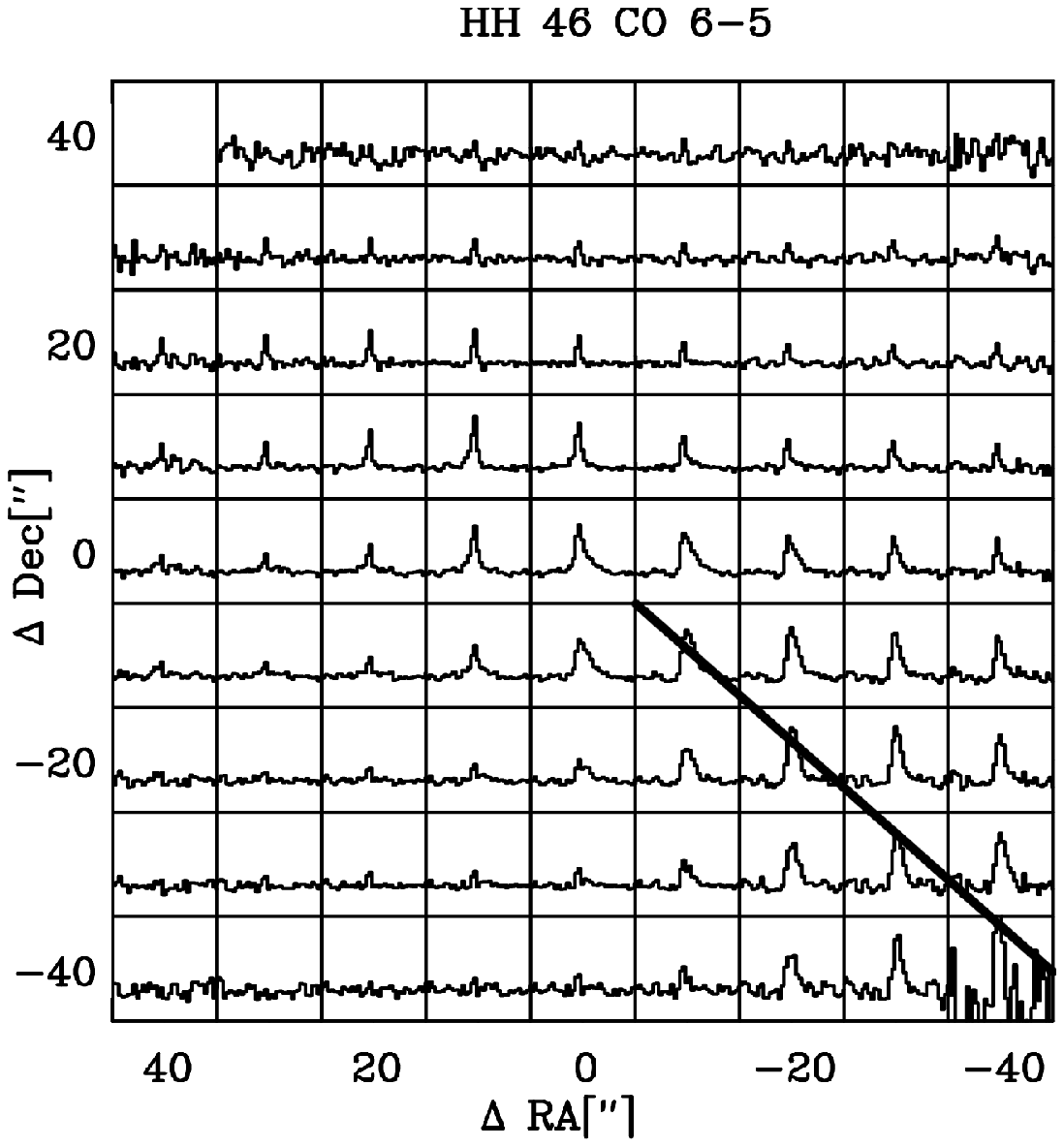}
\end{center}
\caption{Spectra of HH~46 of CO 3--2 ({\it left}) and CO 6--5 ({\it
right}) in the 80$''\times80''$ mapping area. Individual spectra are
shown on a $T_{\rm{MB}}$ scale of -5 to 20 K for CO 3--2 and from -3
to 12 K for CO 6--5. Both axes are from -10 to 20 km s$^{-1}$. The black lines show the axis of the red outflow.}
\label{5:fig:mapsspec}
\end{figure*}
}

\def\placeFigureChapterFiveSix{
\begin{figure}[th]
\begin{center}
\includegraphics[width=170pt]{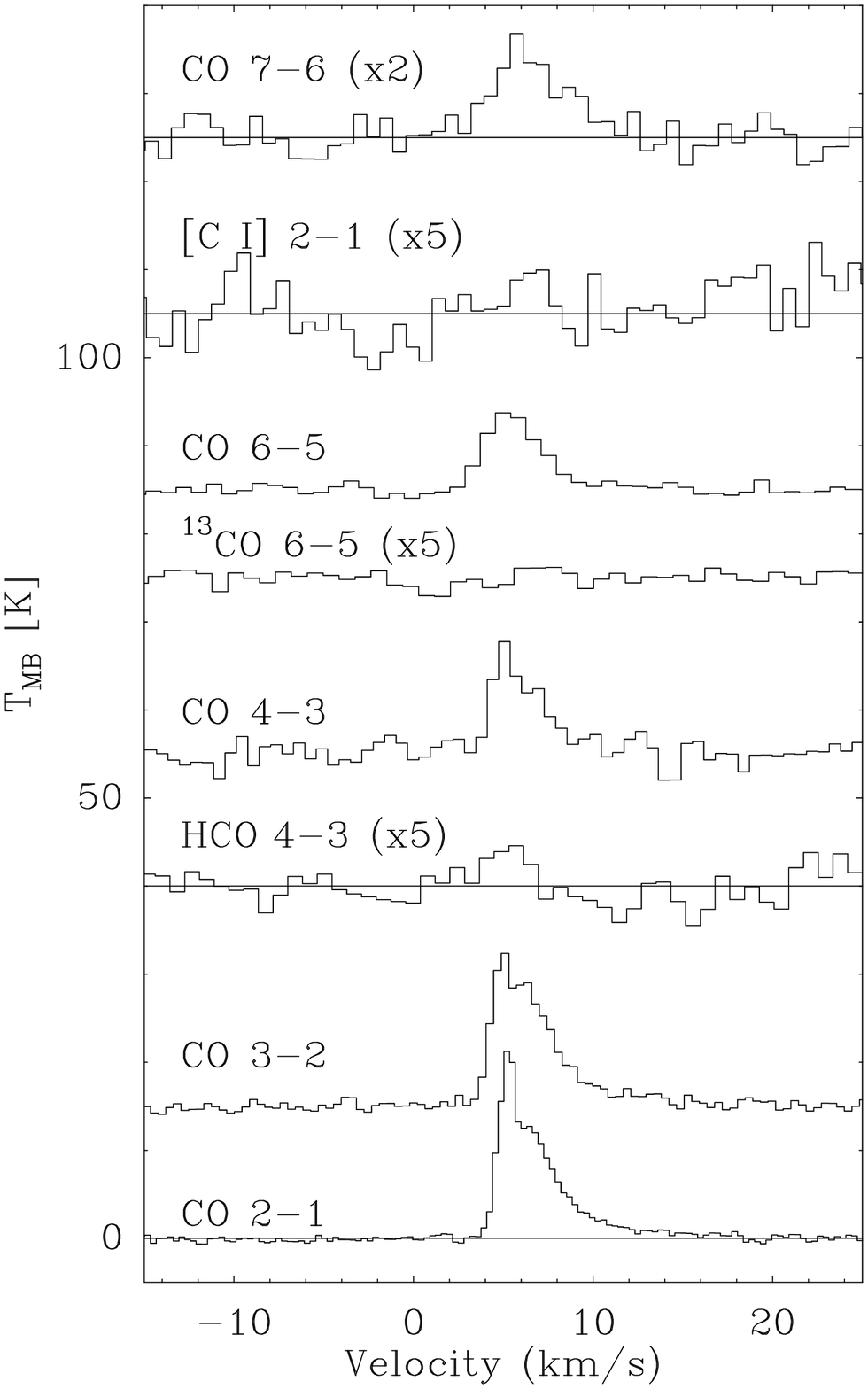}
\end{center}
\caption{Single spectra of (from {\it bottom} to {\it top}): CO 2--1,
CO 3--2, HCO$^+$ 4--3, CO 4--3, $^{13}$CO 6--5, CO 6--5, [C I] 2--1
and CO 7--6. All spectra are at an position in the red outflow lobe
of $\Delta$RA=-20$''$ and $\Delta$Dec=-20$''$. The quiescent gas has a FWHM of 1.5 km s$^{-1}$ centered at 5.3 km s$^{-1}$. }
\label{5:fig:off}
\end{figure}
}

\def\placeFigureChapterFiveSeven{
\begin{figure}[th]
\begin{center}
\includegraphics[width=240pt]{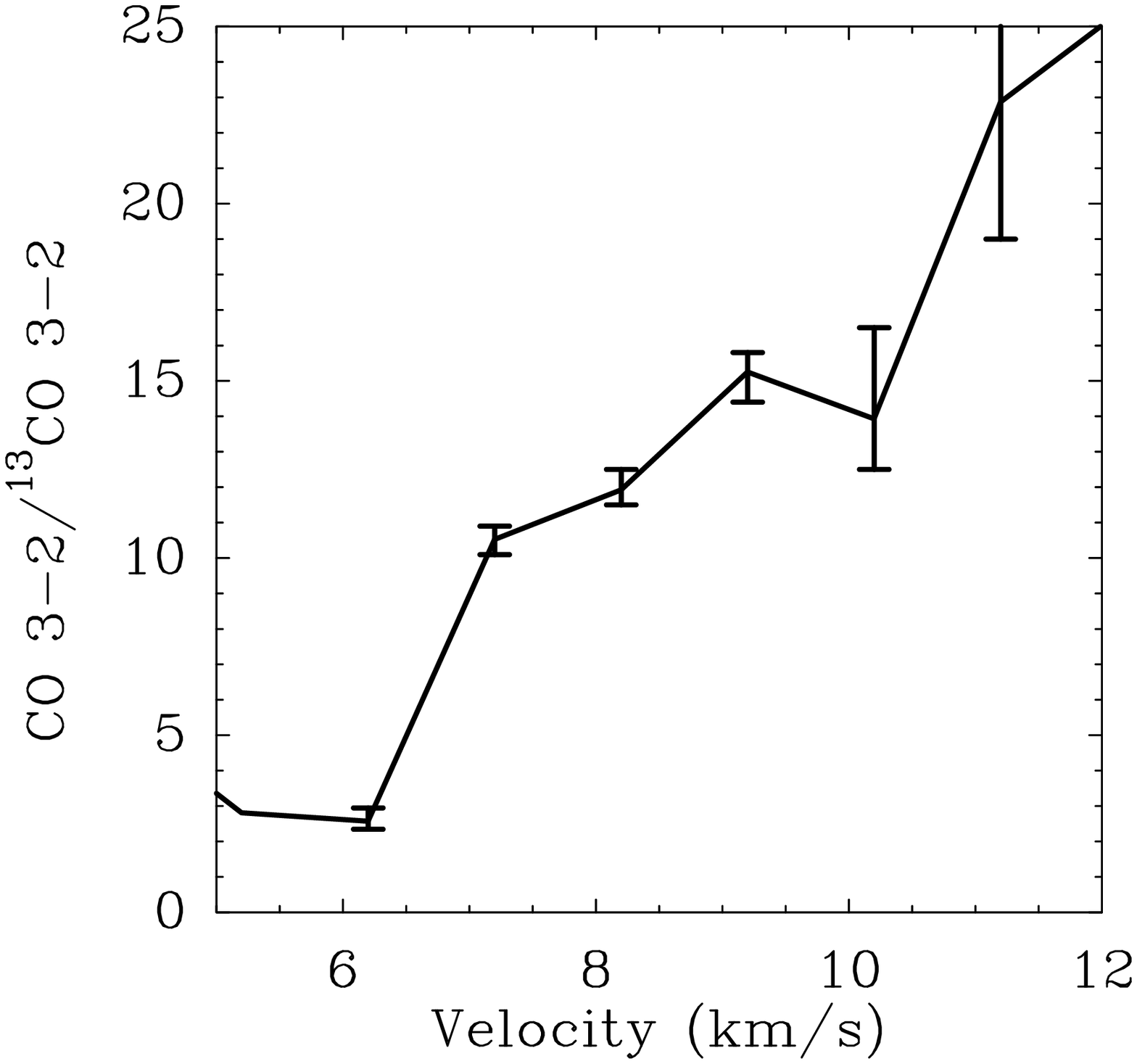}
\end{center}
\caption{The ratio of the main beam antenna temperatures of $^{12}$CO
with respect to $^{13}$CO for the $J=$3--2 line at the (0,0)
position. The ratios correspond to optical depths of 1.8 (ratio of 10)
to 1.0 (ratio of 25). }
\label{5:fig:CO_optdepth}
\end{figure}
}

\def\placeFigureChapterFiveEight{
\begin{figure}[th]
\begin{center}
\includegraphics[width=250pt]{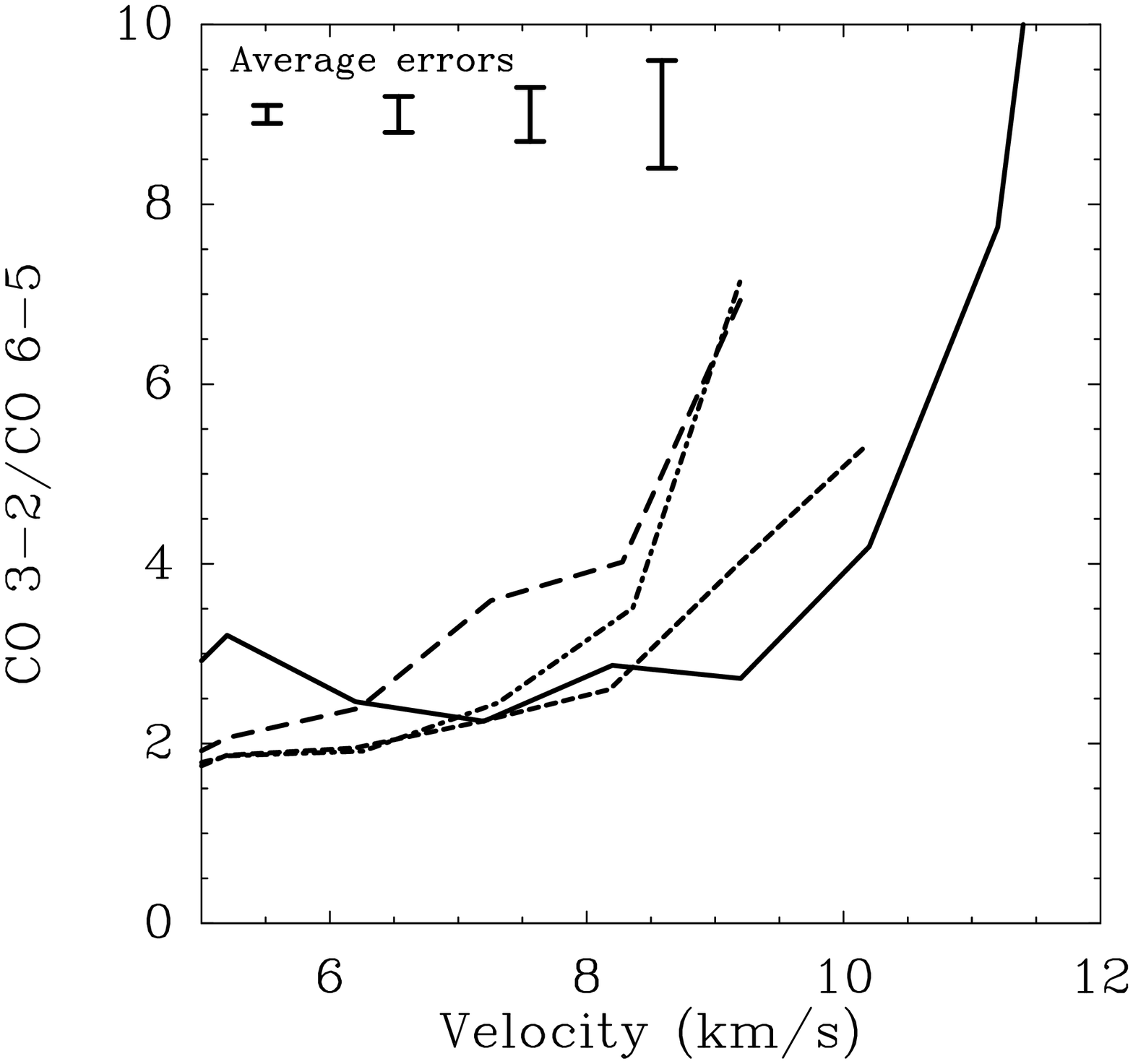}     
\end{center}
\caption{The ratio of the $^{12}$CO 3--2/$^{12}$CO 6--5 main beam
antenna temperatures at various positions along the red outflow axis
({\it solid:}(0,0), {\it dashed:}(-20$''$,-20$''$), {\it
dot-dashed:}(-30$''$,-30$''$), {\it long dash:}(-40$''$,-35$''$)) as
functions of velocity. Average error bars at various velocities
are given in the upper part of the figure at their respective velocity. They are applicable for all positions except
(0,0), for which errors are a factor of 2 lower due to
the longer integration time of CO 3--2 at the center position.}
\label{5:fig:co_ratio}
\end{figure}
}

\def\placeFigureChapterFiveNine{
\begin{figure}[th]
\begin{center}
\includegraphics[width=250pt]{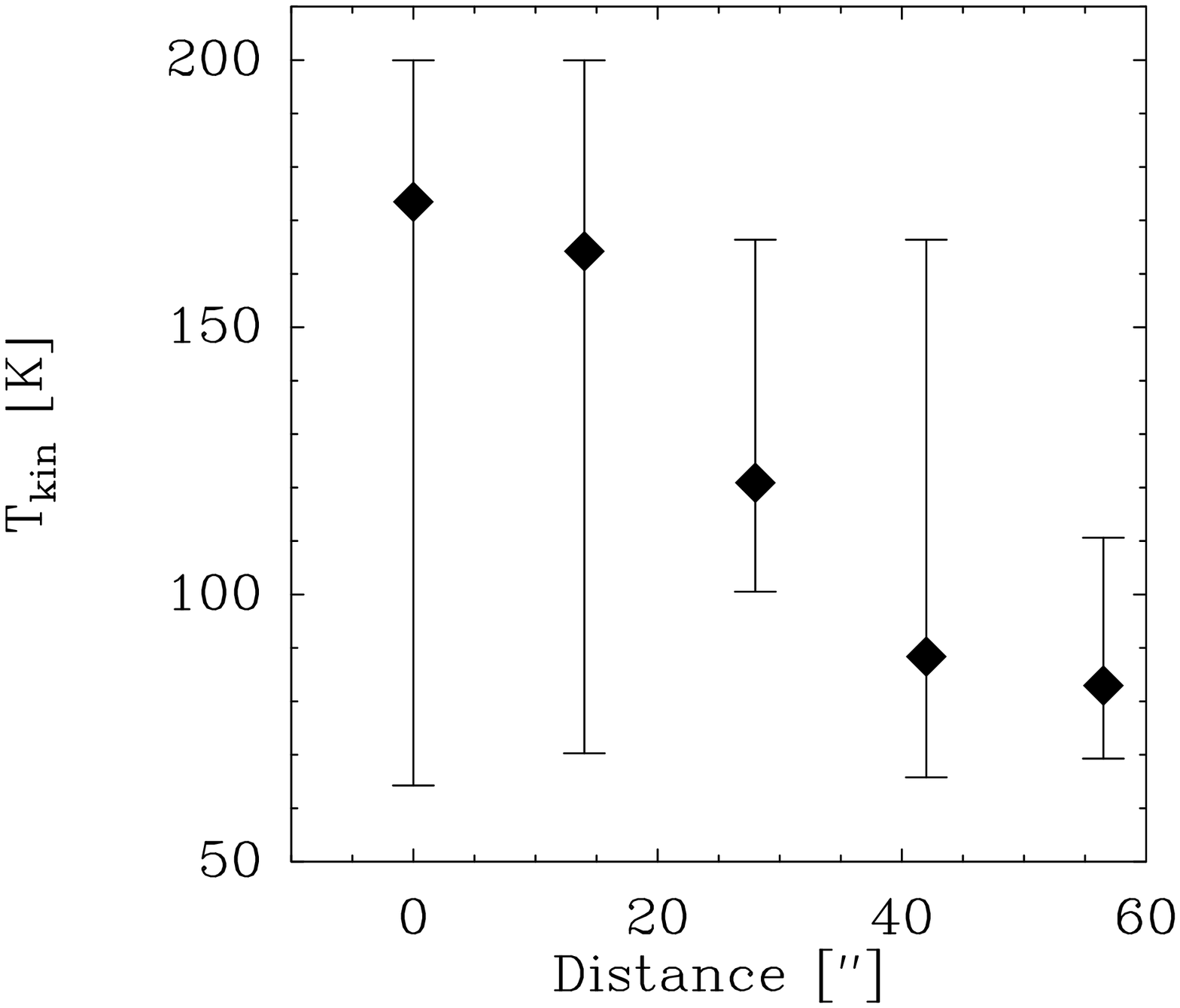}     
\end{center}
\caption{The kinetic temperature of the outflow gas along the red
outflow axis computed from the CO 3--2/6--5 line ratio for a constant
density of 2$\times10^4$ cm$^{-3}$. Diamonds are the temperatures
derived from the velocity-averaged ratios. The limits are determined
from the minimum and maximum ratios between 7 and 11 km s$^{-1}$. 
The results are consistent with a constant temperature of $\sim$100 K if
variations in optical depth and density with distance are taken into account.}
\label{5:fig:co_temp}
\end{figure}

}

\def\placeFigureChapterFivetotal{
\begin{figure*}[th]
\begin{center}
\includegraphics[angle=270,width=500pt]{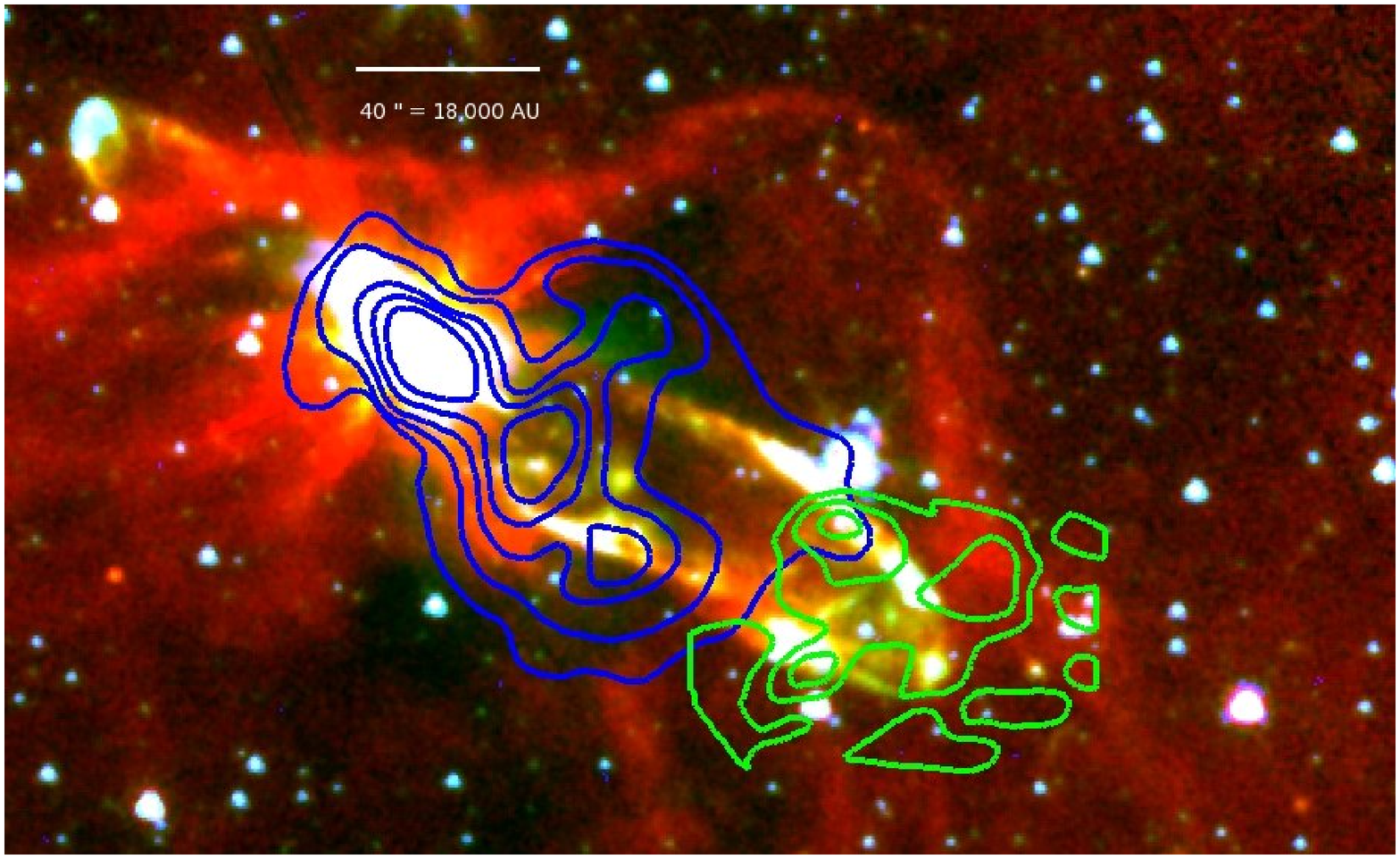}
\end{center}
\caption{The integrated intensity of CO 6--5 (blue contours) and [C I] 2--1 (green contours) overplotted on the Spitzer-IRAC 1 (blue), 2 (green) and 4 (red) bands of the entire HH46 region. CO 6--5 contours are in increasing order of 5 K km s$^{-1}$. The clear cut between the CO and [C I] emission near the bow shock suggests that UV photons capable of dissociating CO are present near the bow shock. }
\label{5:total}
\end{figure*}
}

\def\placeFigureChapterFiveTen{
\begin{figure}[th]
\begin{center}
\includegraphics[width=250pt]{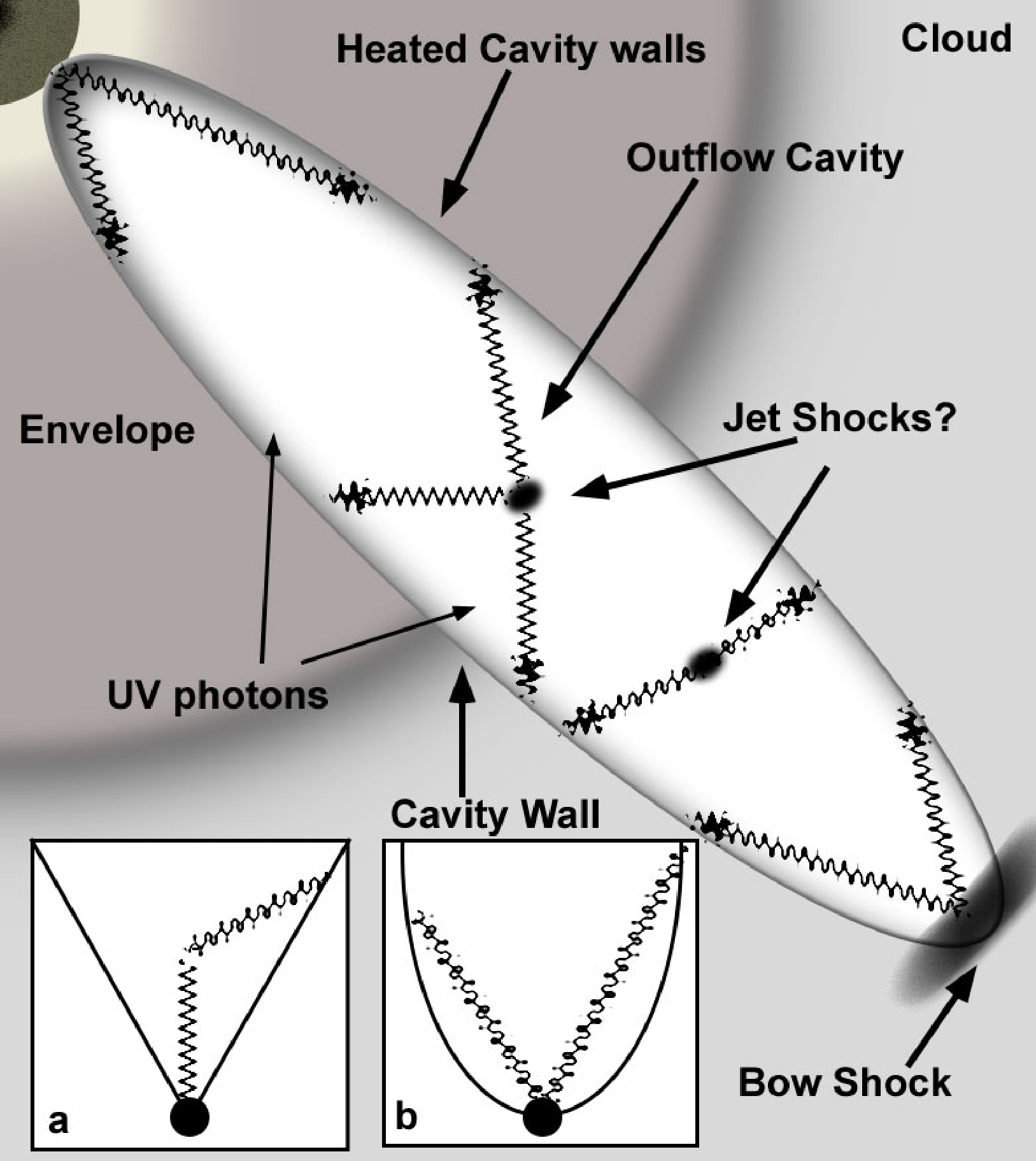}
\end{center}
\caption{Cartoon model of the HH~46 outflow on scales up to the bow shock ($\sim$60,000 AU$\times$$\sim$60,000 AU), illustrating the photon
heating of the cavity walls that is responsible for the high-$J$ CO
emission.  UV photons are created both in the accretion disk boundary
layer, as well as the bow shock and illuminate the cavity walls. Illustrative paths of the UV photons are shown with
wiggly lines. In inset {\bf a}, the method of photon heating proposed by \citet{Spaans95} using an outflow with constant opening angle is shown. There, dust must scatter UV photons to produce significant irradiation of the cavity walls. In inset {\bf b}, a parabolic shape of the outflow cavity allows many more UV photons to directly impact the outflow cavity walls. Due to the expected long mean free path of UV photons within the outflow cavity ($\sim$65,000 AU), scenario {\bf b} is assumed to be the more likely scenario. See the text for more information. }
\label{5:cartoon}
\end{figure}
}

\def\placeFigureDens{
\begin{figure}[!th]
\begin{center}
\includegraphics[width=250pt]{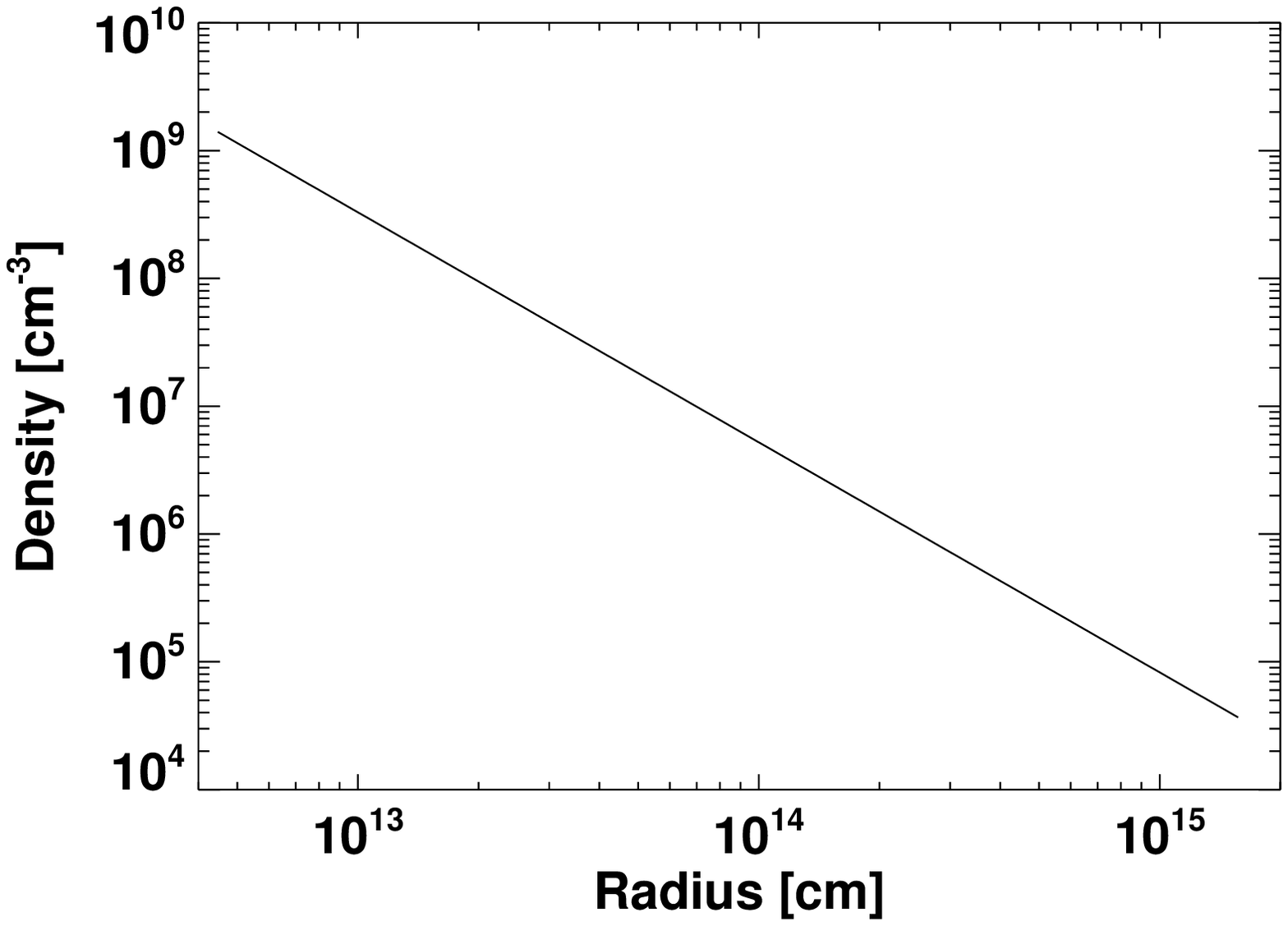}
\includegraphics[width=250pt]{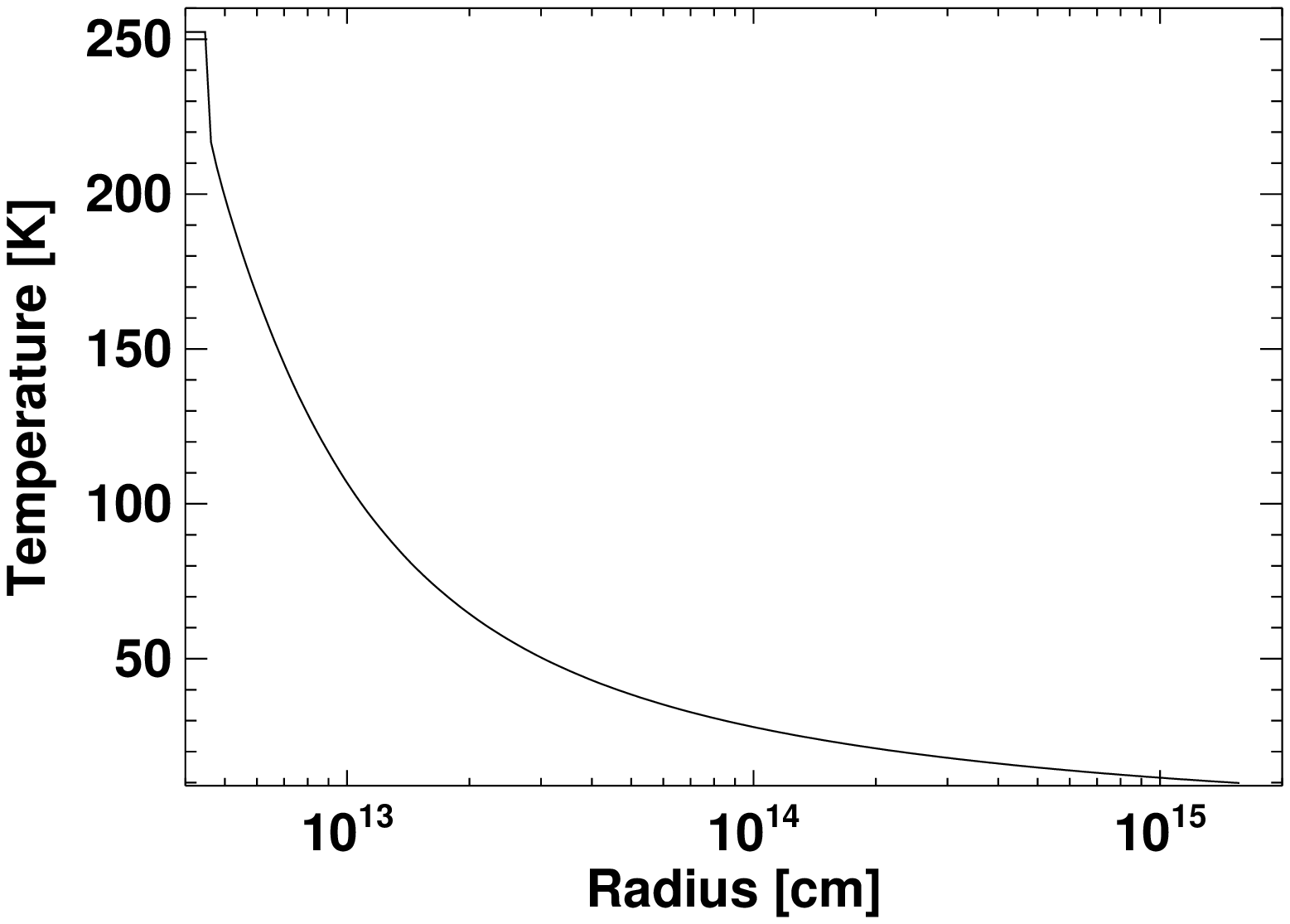}
\end{center}
\caption{The density (\textit{left}) and temperature (\textit{right}) distributions of the DUSTY modelling of HH~46.}
\label{5:dens}
\end{figure}
}

\def\placeFigureRebin{
\begin{figure}[!th]
\begin{center}
\includegraphics[width=200pt]{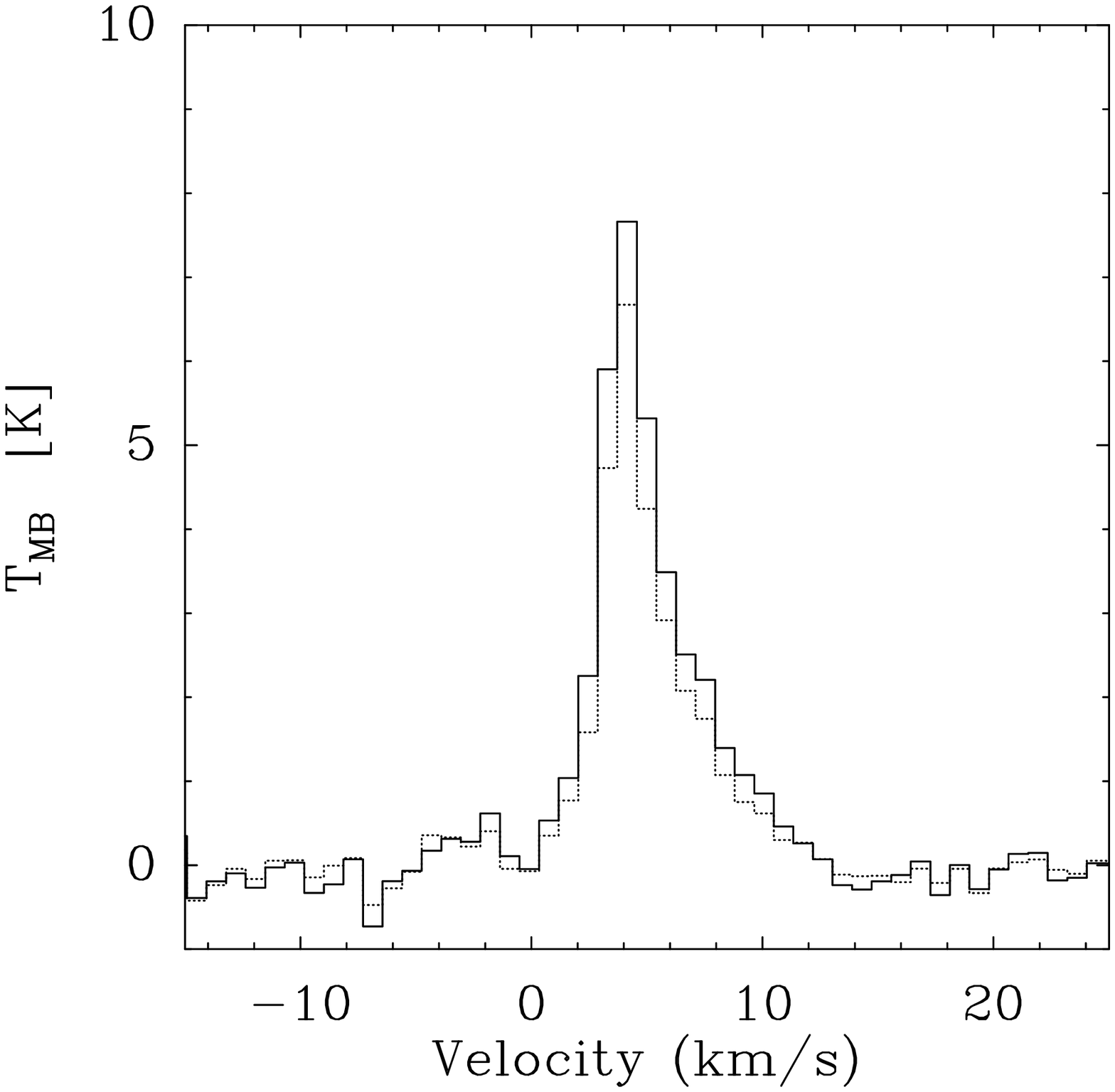}
\end{center}
\caption{The CO 6--5 spectrum at the 0,0 position rebinned at 10$''$ beam (solid line)  and with 20$''$ (dotted line). }
\label{5:rebin}
\end{figure}
}



\def\placeTableChapterFiveOne{
\begin{table*}[th]
\caption{Overview of the observations}
\tiny
\begin{center}
\begin{tabular}{l l l l l r l l}
\hline \hline
Mol. & Trans.& $E_u$ & Freq. & Instrument & Beam & Map & Map Size\\ 
 & $J_{\rm{u}}$-$J_{\rm{l}}$ & (K) & (GHz) &  &($''$)& \\ \hline
CO & 2--1 & 16.6 &230.538 & APEX-1 & 28 & y & 200$"\times170"$ \\
   & 3--2 & 33.2 &345.796 & APEX-2a & 18&  y & 80$"\times80"$\\
   & 4--3 & 55.3 & 461.041 & FLASH-I & 14 &y & 40$"\times40"$ \\
   & 6--5 & 116.2 & 691.473 & CHAMP$^+$-I & 9 &y & 80$"\times80"$$^a$\\
   & 7--6 & 154.8 & 806.651 & CHAMP$^+$-II & 8 &y &80$"\times80"$\\
 $^{13}$CO & 3--2 & 31.7 &330.588 & APEX-2a & 18 & n & -\\
  & 4--3 & 52.9 & 440.765 &FLASH-I & 14 &n  & -\\
  & 6--5 & 111.1 & 661.067 &CHAMP$^+$-I & 9 & y & 50$"\times50"$\\
  & 8--7 & 190.4 & 881.273 &CHAMP$^+$-II & 8 & n & - \\
 C$^{18}$O & 3--2 & 31.6 & 329.331 &APEX-2a & 18 &n & -\\
 & 6--5 & 110.6 & 658.553 & CHAMP$^+$-I & 9 &n & - \\
$[$C I$]$ & 2--1 &  62.3 & 809.342 & CHAMP$^+$-II & 8 & y & 50$"\times50"$$^a$\\
HCO$^+$ & 4--3 & 42.8 & 356.724 & APEX-2a& 18 & y &80$"\times80"$\\\
H$^{13}$CO$^+$ & 4--3 & 41.6 &346.998 & APEX-2a  & 18 &n & -\\ \hline
\multicolumn{4}{c}{870 $\mu$m continuum} & LABOCA & 18 & y & 6$'\times6'$ \\ \hline
\end{tabular}
\end{center}
$^a$ additional mapping was performed in these lines around the bow shock position.
\label{5:tab:lines}
\end{table*}
}

\def\placeTableChapterFiveTwo{
\begin{table*}[th]
\caption{Observed molecular line intenstities$^a$. }
\small
\begin{center}
\begin{tabular}{l l l l l l l l}
\hline \hline
Line & Transition & \multicolumn{4}{c}{Central position} &\multicolumn{2}{c}{Envelope/outflow position$^b$} \\
& & $\int T_{\rm{MB}} \rm{d}V$$^c$ & $T_{\rm{peak}}$ & Red & Blue& $\int T_{\rm{MB}} \rm{d}V$$^c$& $T_{\rm{peak}}$  \\
 &  & (K km s$^{-1}$) & (K) &  (K km s$^{-1}$) &  (K km s$^{-1}$) & (K km s$^{-1}$) & (K)  \\ \hline
CO & 2--1 & 53.3 & 22.4 & 11.5   & 3.8 & 61.7 & 21.2 \\
 & 3--2   & 82.5 & 19.4 & 28.0  & 13.4& 65.6 & 19.8\\
  & 4--3 & 70.5& 14.9 & 44.2 & 3.7 & 38.5 & 14.0\\
 & 6--5 & 42.6 & 9.5 & 11.7 & 5.7 & 34.6  & 8.8\\
 & 7--6 & 46.5 & 8.6 & 17.6 & $<$3.7 & 25.3 & 5.9  \\
 $^{13}$CO & 3--2 & 14.3 & 11.4 & - & - & - & -\\
& 4--3 & 8.0 & 5.5 & - & - & - & - \\
 & 6--5 & 3.9 & 2.9 & -&- & $<$0.6$^d$ & - \\
 & 8--7 & - & $<$0.9$^c$ & -&- & - & - \\
 C$^{18}$O & 3--2 & 3.2 & 3.3 & - & - & - & -\\
          & 6--5 & - & $<$ 0.5$^d$&- &- & - & -\\
$[$C I$]$ & 2--1 & 2.3 & 2.7 &- & - & - & $<$1.8$^d$ \\
HCO$^+$ & 4--3 & 8.5 & 5.4 & 1.4 & - & - & $<$0.9$^d$ \\
H$^{13}$CO$^+$ & 4--3 & 1.1 & 1.0  & -& - & - & -\\ \hline 
\end{tabular}\\
\end{center}
$^a$ Red and Blue shifted emission are calculated by subtracting the central part of the line profile with a gaussian with FWHM of 1.5-2 km s$^{-1}$.\\
$^b$ Position : $\Delta$RA = $-20''$, $\Delta$Dec = $-20''$. \\
$^c$ Velocity range used for integration: -5 km s$^{-1}$ to 15 km s$^{-1}$\\
$^d$ Upper limit of 3$\sigma$ in a 0.7 km s$^{-1}$ channel.\\

\label{5:tab:line_res}
\end{table*}
}

\def\placeTableChapterFiveThree{
\begin{table}[th]
\caption{Results from radiative transfer modelling for envelope properties}
\small
\begin{center}
\begin{tabular}{l l l l}
\hline \hline
\multicolumn{2}{c}{Best fit} & \multicolumn{2}{c}{Envelope Properties} \\ \hline
{\it Y} & 700 & $R_{\rm{inner}}$(250 K) & 30 AU\\
$p$ &  1.8 & $R$(30 K) & 533 AU\\
$\tau_{100}$ & 2.6 &$R$($n_{\rm{de}}$) & 6,100 AU \\
& &  $R_{\rm{outer}}$ & 20,800 AU\\ 
 & & $n$(1000 AU) & 2.5$\times10^{6}$ cm$^{-3}$\\
& &  $M_{\rm{env}}$ ($>$ 10 K) & 5.1 M$_{\rm{\odot}}$\\ \hline
\end{tabular}\\
\label{5:table:prop}
\end{center}
\end{table}
}

\def\placeTableChapterFiveFour{
\begin{table}[!th]
\caption{Results from molecular line radiative transfer modelling.}
\small
\begin{center}
\begin{tabular}{l l l l l l}
\hline \hline
Line & \multicolumn{2}{c}{Model 1$^a$} &\multicolumn{2}{c}{Model 2$^a$} & Obs.\\ 
 & $\int T_{\rm{MB}}$ & $T_{\rm{peak}}$$^b$ & $\int T_{\rm{MB}}$ & $T_{\rm{peak}}$$^b$ & $T_{\rm{peak}}$  \\ \hline
CO 2--1 & 9.4 & 5.0/22.2 & 10.4 & 4.5/78.9 & 22.4\\
CO 3--2 & 8.0 & 4.7/22.5 & 7.1  & 3.8/55.8 & 19.4\\
CO 4--3 & 7.3 & 4.2/16.9 & 4.7  & 3.1/47.3 & 14.9\\
CO 6--5 & 6.8 & 3.4/4.8 & 5.1  & 2.7/5.2 & 9.5\\
CO 7--6 & 4.1 & 2.2 & 4.0 & 2.2  & 8.6\\
$^{13}$CO 3--2 & 4.6 & 3.7/3.8 & 5.9 & 3.6/10.7 & 11.4\\
$^{13}$CO 4--3 & 3.5 & 2.8 & 4.2 & 2.6/3.8 & 5.5\\
$^{13}$CO 6--5 & 1.9 & 1.2 & 1.9 & 1.2 & 2.9\\
$^{13}$CO 8--7 & 0.8 & 0.5 & 0.8 & 0.5 &$<$0.9\\
C$^{18}$O 3--2 & 1.3 & 1.2 & 3.5 & 2.9 & 3.3\\
C$^{18}$O 6--5 & 1.0 & 0.7 & 1.0 & 0.7 & $<$0.5\\ \hline
HCO$^+$$^c$ 4--3 & 3.5 & 2.3/11.8 & 2.8 & 2.5/8.3 & 5.4\\ 
H$^{13}$CO$^+$ 4--3 & 0.75 & 0.7 & 0.75 & 0.7 & 1.0\\ \hline
\end{tabular}\\
\label{5:tab:coprop}
\end{center}
$^a$ Model 1 has a jump abundance profile with $X_0$/$X_d$ of
2.7$\times10^{-4}$/1$\times10^{-5}$. Model 2 has a drop abundance
profile with $X_0$/$X_d$/$X_0$ of
2.7$\times10^{-4}$/1$\times10^{-5}$/2.7$\times10^{-4}$.\\ 
$^b$ For
some lines, the peak temperature is given as X/Y. The first value is
the actual $T_{\rm{peak}}$ in the model profile. The second value
refers to the peak of a gaussian fitted to the line wings of model
profiles which show self-absorption.  If only a single number is
given, the modelled line is gaussian in nature and does not have
self-absorption. \\
$^c$ HCO$^+$ abundances of $X_0$ and $X_d$ are 2.0$\times$10$^{-8}$ and 3.0$\times$10$^{-9}$. 
\end{table} 
}

\def\placeTableChapterFiveFive{
\begin{table*}[th]
\caption{Outflow properties of the red and blue outflow lobe). }
\small
\begin{center}
\begin{tabular}{l l l l l l l l}
\hline \hline
\multicolumn{8}{c}{Outflow properties} \\ \hline
& $V_{\rm{max}}$$^a$ & $R$ &$M$$^b$$^{,c}$ & $t_{\rm{d}}$$^a$$^{,d}$ & $\dot{M}$$^a$$^{,e}$ & $F_{CO}$$^b$$^{,f}$ & $L_{\rm{kin}}$$^{b,g}$  \\ %
 & (km s$^{-1}$) & (AU)& (M$_\odot$) & (yr) & (M$_\odot$  &(M$_\odot$ yr$^{-1}$ & (L$_\odot$)  \\ %
             &     &           &     &      &  yr$^{-1}$)& km s$^{-1}$) & \\\hline %
 \multicolumn{8}{c}{Red Lobe} \\ \hline
CO 2--1 & 14.2 &  5e4      & 0.8/2.9$^h$     &  2.6e4   & 9.1e-5  & 9.8e-4  & 1.2e-2 \\
CO 3--2 & 14.0 & $>$2.5e4  & 0.68    &$>$1.3e4  & 4.2e-5  & 4.5e-4  & 5.4e-3\\
CO 4--3$^i$ & 11.0 & -     & $>$0.18 & -         & -    & - & -\\
CO 6--5 & 10.5 &  $>$2.5e4 & 0.48    & 2.2e4   & 1.8e-5 & 1.1e-4  & 8.0e-4 \\
CO 7--6 & 12.1 & 2.0e4     & 0.4     & 1.4e4   & 2.2e-5 & 1.7e-4  & 1.6e-3\\ \hline
\multicolumn{7}{c}{Blue Lobe} \\ \hline
CO 2--1 & 0.5 & 1.2e4  & 0.24& 1.2e4  &  1.6e-5  & 9.3e-5  & 6.0e-4 \\
CO 3--2 & 0.3 & 1.6e4  & 0.24& 1.5e4  &  1.3e-5  & 7.6e-5  & 5.0e-4\\
CO 4--3 & 1.7 & 1.1e4  & 0.19& 1.5e4  &  1.1e-5  & 4.5e-5  & 2.2e-4\\
CO 6--5 & 0.5 & 1.1e4  & 7e-3& 1.1e4  &  5.4e-7  & 3.0e-6  & 2.0e-5\\
CO 7--6 & 1.1 & 1.0e4  & 1e-2& 1.2e4  &  8.6e-7  & 4.3e-6  & 2.4e-5\\ \hline
\end{tabular}\\
\end{center}
$^a$ Actual LSR velocities; note that the quiescent gas is at $V_{\rm
LSR}$=5.3 km s$^{-1}$. $\delta V_{\rm{max}}$=abs($V_{\rm{max}}$-$V_{\rm{LSR}}$). Velocities are not corrected for inclination.  \\
$^b$ Corrected for inclination using the average correction factors of
\citet{Cabrit90}\\ 
$^c$ Constant temperature of 100 K assumed for red outflow and 70 K for blue outflow.\\
$^d$ Dynamical time scale :
$t_{\rm{d}}=R/\delta V_{\rm{max}}$\\ 
$^e$ Mass outflow rate :
$\dot{M}=M/t_{\rm{d}}$, not corrected for swept-up gas.\\ 
$^f$ Outflow
force : $F_{CO}=MV_{\rm{max}}^2/R$ \\ 
$^g$ Kinetic luminosity :
$L_{\rm{kin}}=0.5M(\delta V_{\rm{max}})^3/R$\\ 
$^h$ 0.8 M$_\odot$ is derived for a temperature of 50 K. 2.9 M$_\odot$ is derived for a temperature of 100 K \\
$^i$ Outflow extends to far
larger scales than map scale.\\
\label{5:table:out}
\end{table*}
}

  \abstract 
   {The spectacular outflow of HH~46/47 is driven by HH~46 IRS 1, an
   embedded Class I Young Stellar Object (YSO). Although much is known
   about this region from extensive optical and infrared observations,
   the properties of its protostellar envelope and
   molecular outflow are poorly constrained.}
{Our aim is to characterize the size, mass, density and temperature
   profiles of the protostellar envelope of HH~46 IRS 1 and its
   surrounding cloud material as well as the effect the outflow has on
   its environment.}
{The newly commisioned  CHAMP$^+$ and LABOCA arrays on the APEX
telescope, combined with lower frequency line receivers, are used to
obtain a large (5$'\times5'$, 0.6$\times$0.6 pc) continuum map and
smaller (80$''\times80''$, 36,000$\times$36,000 AU) heterodyne maps in
various isotopologues of CO and HCO$^+$. The high-$J$ lines of CO
(6--5 and 7--6) and its isotopologues together with [C I] 2--1, observed with CHAMP$^+$, are
used to probe the warm molecular gas in the inner few hundred AU and
in the outflowing gas. The data are interpreted with continuum and line
radiative transfer models.}
{Broad outflow wings are seen in CO low- and high-$J$ lines at several
positions, constraining the gas temperatures to a constant value of
$\sim$100 K along the red outflow axis and to $\sim$60 K for the blue outflow.
The derived outflow mass is of order 0.4--0.8 M$_\odot$, significantly
higher than previously found.  The bulk of the strong high-$J$ CO line
emission has a surprisingly narrow width, however, even at outflow
positions. These lines cannot be fit by a passively heated model of the HH 46 IRS
envelope. We propose that it originates from photon heating of the
outflow cavity walls by ultraviolet photons originating in outflow
shocks and the accretion disk boundary layers.  At the position of the bow shock itself, the UV photons are energetic enough to dissociate CO. The envelope mass of
$\sim$5 M$_\odot$ is strongly concentrated towards HH 46 IRS with a
density power law of $-1.8$.  }
   {The fast mapping speed offered by CHAMP$^+$ allows the use of
   high-$J$ CO lines and their isotopes to generate new insights into
   the physics of the interplay between the molecular outflow and
   protostellar envelope around low-mass protostars. The UV radiation inferred from the high-J CO and [C I] data will affect the chemistry of other species.}


   \maketitle

\section{Introduction}
The Young Stellar Object (YSO) HH~46 IRS 1 (RA= 08h25m43.9s, Dec
=$-$51d00h36s (J2000) ), located  at the edge of an isolated Bok Globule ($D=$ 450
pc) in the southern hemisphere \citep{Schwartz77}, is well-known for
its spectacular outflow \citep{Reipurth91}, observed at both visible
and infrared (IR) wavelengths with the {\it Hubble} and {\it Spitzer}
Space Telescopes
\citep[e.g.,][]{Heathcote96,Stanke99,NoriegaCrespo04,Velusamy07}. Deep
H$\alpha$ observations using the VLT have revealed bow shocks
associated with the HH 46 outflow up to a parsec away from the central
source \citep{Stanke99}.  Its blue-shifted lobe expands outside the cloud in a low density region, due to the close proximity of the protostar to the edge of the cloud.
 Proper motion and radial velocity studies
show that the outflow has an inclination of 35$^\circ$ with respect to
the plane of the sky and flow velocities in atomic lines up to 300 km
s$^{-1}$ \citep{Dopita82, Reipurth91,Micono98}. The internal driving
source was found to be HH~46 IRS 1 ($L=16$ $L_{\rm\odot}$), an
embedded Class I YSO \citep{Raymond94, Schwartz03}. Surprisingly, much
less is known about the properties of the protostellar envelope and
the molecular outflow.  \citet{Chernin91} and \citet{Olberg92}
mapped this region using low excitation CO lines, which show that
contrary to the optical flows, the red-shifted outflow lobe is much
stronger than the blue-shifted one. \citet{Chernin91} theorised that a
lack of dense material in the path of the blue-shifted flow is responsible
for this.

Comparisons between dust emission and molecular lines at
submillimeter wavelengths, together with self-consistent radiative
transfer calculations, have been extensively used to characterize the
physical and chemical structure of Class 0 and Class I envelopes
\citep{Schoeier02,Jorgensen02, Jorgensen05,Maret04}.  However, an
essential component could not be probed with those data. The amount of
warm ($T>$ 50 K) gas within the protostellar envelope as well as the
influence of the molecular outflow have not been constrained directly
using observations of lower excited molecular lines in the 200 and 300
GHz bands. Although more complex molecules, such as H$_2$CO and
CH$_3$OH emit at these frequencies from high energy levels
\citep[e.g.][]{vanDishoeck95, Ceccarelli00}, their more complex
chemistry complicates their use as probes of the warm
gas. Observations of CO at higher energies (up to 200 K) provide more
reliable probes into the inner regions of envelopes and molecular
outflows, but such lines have only been observed for a handful of
sources \citep[e.g.,][]{Hogerheijde98,Parise06, vanKempen06}.  [C I]
emission provides an important constraint on the strength of the
radiation field within the outflows \citep{Walker93}.

With the commissioning of the {\it Atacama Pathfinder EXperiment}
(APEX)\footnote{This publication is based on data acquired with the
Atacama Pathfinder Experiment (APEX) in programs E-77.C-0217,
X-77.C-0003, X-79.C-0101 and E-081.F-9837A.  APEX is a collaboration
between the Max-Planck-Institut fur Radioastronomie, the European
Southern Observatory, and the Onsala Space Observatory. APEX-1 was
used during science verification in June 2008. } \citep{Guesten06},
the CHAMP$^+$ instrument \citep{Kasemann06,Guesten08} allows observations of
molecular emission lines in the higher frequency sub-millimeter bands
of southern sources, like HH~46.  CHAMP$^+$ is the first array of its
kind. With its 14 pixels, it is able to observe simultaneously in the
690 and 800 GHz atmospheric windows. This combination of
dual-frequency observing and fast mapping speed, supplemented by lower
frequency single pixel data and a LABOCA continuum array map
\citep{Kreysa03,Siringo08}, provides a large range of highly complementary
tracers of both the gas and dust conditions in the inner and outer
regions of the envelope, as well as the molecular outflow on scales of
a few arcminutes. Spectral line maps provide key information that is
essential in the analysis of embedded YSOs which single-pointed
observations cannot offer \citep{Boogert02,vanKempen08}.

In this paper, we present first results from CHAMP$^+$ and LABOCA
observations of the HH 46 IRS 1 source, supplemented by lower
frequency line receivers. Observations of highly excited CO, HCO$^+$
and their isotopologues are used to constrain the properties of the
protostellar envelope and molecular outflow of HH~46.  In $\S$ 2 the
observations performed at APEX are presented. Resulting spectra and
maps can be found in $\S$ 3. In $\S$ 4 and 5 the protostellar envelope,
molecular outflow and close surrounding of HH~46 are characterized
using a radiative transfer analysis. In addition, we discuss a
possible scenario for the high-$J$ CO emission in $\S$ 6. The final conclusions
are given in $\S$ 7. CHAMP$^+$ observations of a larger sample of low-mass protostars are presented in a subsequent paper (van Kempen et al. submitted).

\placeTableChapterFiveOne

\section{Observations}

Molecular line observations were carried out with the CHAMP$^+$ array
\citep{Kasemann06} of CO and its isotopologues, ranging in transitions
from $J$=6--5 to $J$=8--7, as well as [C I] 2--1. These observations
were supplemented with low excitation line observations using APEX-1
(230 GHz, CO 2--1), APEX-2a (345 GHz, CO, C$^{18}$O and $^{13}$CO
3--2, and HCO$^+$ and H$^{13}$CO$^+$ 4--3) and FLASH (460 GHz, CO and
$^{13}$CO 4--3). In addition, LABOCA was used to map the entire region
at 870 $\mu$m.  See Table \ref{5:tab:lines} for an overview of the
observed emission lines for each instrument and the corresponding rest
frequencies and upper level energies, together with their
corresponding APEX beams. All observations were done under excellent
weather conditions with typical system temperatures of 2100 K for
CHAMP$^+$-I (SSB, 690 GHz), 7500 K for CHAMP$^+$-II (SSB, 800 GHz), 1100 K for
FLASH-I (DSB, 460 GHz), 290 K for APEX-1 and 230 K for APEX-2a (both SSB). Calibration errors are estimatd at 15 to 20 $\%$. The HH~46
protostar was spectrally mapped in CO 2--1, 3--2, 4--3, 6--5, 7--6 and
HCO$^+$ 4--3, as well as $^{13}$CO 6--5 and [C I] 2--1. The mapped
area differs per line, ranging from $40''\times 40''$ for CO 4--3 to
$200''\times 200''$ for CO 2-1, with most other lines covering
$80''\times 80''$. Observations were taken over a period of 2 years
from July 2006 to September 2008 using Fast Fourier Transform Spectrometer
(FFTS) \citep{Klein06} backends for all instruments, except CHAMP$^+$,
for which only the two central pixels were attached to the FFTS
backends. Other CHAMP$^+$ pixels were attached to the MPI Array
Correlator System (MACS) backends. FFTS backends are able to reach
resolutions of 0.12 MHz (0.045 km s$^{-1}$ at 800 GHz), while the MACS
units were used at a resolution of 1 MHz (0.36 km s$^{-1}$ at 800
GHz).  Beam efficiencies are 0.75 for APEX-1, 0.73 for APEX-2a, 0.7
for FLASH-I, 0.56 for CHAMP$^+$-I and 0.43 for CHAMP$^+$-II. Pointing
was checked on nearby sources and found to be accurate within 3$''$ for the APEX-2a
observations. For CHAMP$^+$, pointing is accurate within
$\sim$5$''$. All observations were taken using position switching with
reference positions in azimuth ranging from 600$''$ to 3600$''$.

LABOCA observed HH~46 during November 2007 using its 295 pixels in a
 spiral mode centered on the IR position using a position switch of
 600$''$ arcseconds. Only the inner 5$'\times5'$ of the 11$'$ field of
 view was used. The field was integrated down to a noise level of 0.01
 Jy/beam, averaged over the entire region. Due to the used spiral
 mode, the map contains a radial increase of noise towards the edge of
 the map. The continuum data were reduced with the BOA
 package\footnote{http://www.astro.uni-bonn.de/boawiki}.

\placeFigureChapterFiveOne  

\placeTableChapterFiveTwo
\placeFigureChapterFiveTwo
\placeFigureChapterFiveThree

\section{Results}

\subsection{Single pixel spectra}
Figure \ref{5:fig:spec_2a} shows the spectra taken at the position of HH
46 IRS 1 for all lines in Table \ref{5:tab:lines}. Integrated
intensities, peak temperatures and estimated contributions from the
blue- and red-shifted outflowing gas are given in Table
\ref{5:tab:line_res}. The latter are derived by subtracting the central
part of the line profile, associated with the quiescent gas, with a
single gaussian. Emission was detected for all lines with the
exception of C$^{18}$O 6--5 and $^{13}$CO 8--7. The quiescent gas
component peaks at a $V_{\rm{LSR}}$ of 5.3 $\pm$0.1 km s$^{-1}$ and
has a FWHM of 1-2 km s$^{-1}$ depending on the energy of the upper
level. Since the detection of the [C I] 2--1 line is only 3.5
$\sigma$, a gaussian fit is overplotted in Fig. \ref{5:fig:spec_2a}.

Integrated intensities range from 82.5 K km s$^{-1}$ for CO 3--2 to
1.1 K km s$^{-1}$ for H$^{13}$CO$^+$. All $^{12}$CO line profiles show
contributions of a red-shifted outflow lobe within the beam, including
the high excitation CO 7--6 line. Emission from the blue-shifted
outflow lobe is much weaker and not found for CO 7--6. In the other
lines, outflow emission is only detected for HCO$^+$ 4--3, where a
weak red-shifted wing is found.

\subsection{Maps around IRS 1}
The resulting dust emission map (5$'\times5'$) at 870 $\mu$m is shown
in Fig. \ref{5:fig:laboca}. The envelope is clearly resolved,
considering the beamsize of 18$''$ of the APEX dish. The envelope is
slightly elongated on a south-west to north-east axis. South of HH~46,
dust emission from the cold cloud is seen. The total integrated flux
in a 120$''$ diameter aperture around the source position is found to
be 3.7 Jy. Reduction of archival SCUBA (850 $\mu$m) data of this
region yielded a flux of 3.3 Jy \citep{diFrancesco08}, a difference in
flux well within the estimated calibration errors of both SCUBA and
LABOCA. Using the formula in \citet{Shirley00}, the total mass in the
300$''$ mapped area of the cloud is 8 M$_\odot$,  while the central 120$''$ is associated with a mass of 3.7 $M_\odot$, both with an assumed dust temperature of 20 K.

The integrated spectral line maps of CO 3--2, 7--6 and HCO$^+$ 4--3
(all 80$''\times80''$) in Fig. \ref{5:fig:maps2a} show a similar
structure as the continuum maps (see also $\S$5.4.1), although the elongation to the
south-west is much more pronounced in the CO lines, with an integrated
intensity of CO 3--2 of 65.6 K km s$^{-1}$ seen in the -20$''$,-20$''$
position. As can be seen from the red- and blue-shifted velocity
maps of CO 2--1, 4--3 and 6--5 in Fig. \ref{5:fig:mapsvel}, the shape of
the integrated intensity maps is largely due to the outflow
contributions, especially for the low-excitation lines, even in the CO
2--1 map, for which a map of 200$''\times200''$ was observed. This is
in agreement with the results from \citet{Chernin91}. In fact, the CO
3--2 line is so prominent that it can contribute significantly to the
`continuum' emission seen by LABOCA. The CO 3--2 line produces a flux
density of 40 mJy/beam off source, equivalent to the LABOCA 4$\sigma$
level. The total continuum emission seen at these positions is
$\sim$12--15$\sigma$. CO emission may thus contribute up to 30$\%$ to
the observed LABOCA emission in the outflow region.  Another possible contribution to the dust emission at 870 $\mu$m is heating of the dust grains by UV radiation that takes place in the cavity walls.  This is further discussed in $\S$ 6.

In contrast, the elongation in the HCO$^+$ 4--3 map is much less
pronounced than in the CO low-$J$ data. 
The map of $[$C I$]$ 2--1 is not shown, since no lines were detected
down to 1.8 K (3$\sigma$), except at the source position (Fig. 1) and at the bow shock (see $\S$ 3.3). In the map of
$^{13}$CO 6--5, lines are only detected at the central position and at
neighbouring pixels along the outflow.

\placeFigureChapterFiveFour
\placeFigureChapterFivetotal
\placeFigureChapterFiveFive

Fig. \ref{5:total} shows the total distribution of $^{12}$CO 6--5 over the entire area, overplotted over the Spitzer image of \citet{NoriegaCrespo04} with the Spitzer-IRAC 1,2 and 4 bands (3.6, 4.5 and 8.0 $\mu$m respectively.) The CO 6--5 integrated intensities follow the outflow but do not extend all the way out to the bow shock. In addition, several maximum intensities can be found. The first one corresponds with the protostellar envelope, but the other two seem to be related to the outflow. Similarly regularly spaced 'knots' are seen in , for example, the CO 3--2 maps of the NGC 1333 IRAS4A outflow \citep{Blake95}.

 Figure \ref{5:fig:mapsspec} shows the observed spectra within the
 central 80$''\times80''$ of CO 3--2, taken with APEX-2a, and CO 6--5,
 taken with CHAMP$^+$-I. Both are binned to square 10$''\times10''$
 pixels. The line profiles show an interesting distribution of
 emission, especially at the positions associated with the outflowing
 gas. The CO 3--2 spectra consist of a central gaussian originating in
 the cold envelope material surrounding the protostar, with a strong
 red-shifted outflow of up to 10 km s$^{-1}$ away from the source and
 cloud velocity in the south-east direction. The line profile at the
 source position has a quiescent component, flanked by both red- and
 blue-shifted emission. Interestingly, the CO 6--5 emission shows a
 quiescent component with relatively weak red-shifted emission at other positions in the map. The central part of the line profile can be fitted
 with narrow ($\Delta V$ = 2-3 km s$^{-1}$) gaussian profiles. At the
 central position, outflow emission in the CO 6--5 line is much more
 prominent, but a strong quiescent component is still present.
\placeFigureChapterFiveSix

Figure \ref{5:fig:off} shows all spectra observed within the spectral
line maps at a relative position of $(-20''$,$-20'')$. Peak temperatures and
integrated intensities are given in Table \ref{5:tab:line_res}. This
position covers the red-shifted outflow seen prominently in the
emission of the low-excitation CO 3--2 and 2--1 lines.
Even CO 4--3 shows significant red-shifted outflow emission. The
outflowing gas is still present in the high-$J$ CO lines. However, for
both CO 6--5 and 7--6 a significant part of the emission (on the order
of 70-80$\%$) originates in a quiescent narrow component. The isotopologue
$^{13}$CO 6--5 is not detected, down to a
3$\sigma$ level of 0.6 K in a 0.7 km s$^{-1}$ bin. Similarly, no [C I]
is detected down to 1.8 K and no HCO$^+$ 4--3 is detected down to 0.9
K, both 3$\sigma$ in a 0.7 km s$^{-1}$ bin.

 Using the limits on the $^{13}$CO 6--5 emission at both the central
 position and the selected off-postion, it is found that the quiescent
 component at the central position is optically thick ($\tau > 3$),
 while the quiescent emission at the outflow position is optically
 thin ($\tau < 0.4$), as no $^{13}$CO is detected between 0 and 10 km s$^{-1}$.  This analysis
 assumes a $^{12}$CO:$^{13}$CO ratio of 70:1 \citep{Wilson94}.

Although outflow emission heavily influences the line profiles of the
CO 3--2 and 6--5 throughout the maps, positions south, south-east, and
north of the (0,0) position are not affected by any outflowing gas,
as seen in Fig. \ref{5:fig:mapsspec}. About 30$''$ to 40$''$ north of
HH~46, both the CO 3--2 and CO 6--5 are not detected. Even in the map
of CO 2--1, no emission was found at these positions. It is concluded
that the cold cloud material of the surrounding Bok globule does not
extend to these scales. At positions south and south-east of HH~46, CO
3--2 emission is seen, but no CO 6--5 is detected there.

\subsection{Bow shock}
 The bow shock associated with the red outflow lobe is clearly visible in the IR images of \citet{NoriegaCrespo04}, located at a relative offset of (-100$''$,-60$''$) with respect to the source. Additional observations of this bow shock in CO 6--5, [C I] 2--1 and $^{13}$CO 6--5 were carried out over a 90$''$ by 60$''$ area. Fig. \ref{5:total} shows the distribution of [C I] 2--1 emission in this region. Narrow [C I] 2--1 is clearly detected near and at the position of the bow shock, where it has a integrated intensity of 1.2 K km s$^{-1}$ with a peak temperature of 1.7 K. Interestingly, the emission seems to be spatially extended along the outflow axis into the outflow lobe. No $^{13}$CO 6--5 was detected at any position near the bow shock down to an rms of 85 mK in a 0.3 km s$^{-1}$ bin.
The combination of detected [C I] 2--1 and lack of CO 6--5 emission suggests that CO is dissociated by either the shocks present near the bow shock or by UV photons capable of dissociating CO )see $\S$ 6.1).  

\section{Envelope and surrounding cloud}
\subsection{Envelope - dust}
 
\placeFigureDens
Using the 1-D radiative transfer code DUSTY \citep{Ivezic97}, a
spherically symmetric envelope model is constructed by fitting the
radial profile of the 870 $\mu$m image and the SED simultaneously,
determining the size, total mass, inner radius as well as density and
temperature profiles of the protostellar envelope. For a more thorough
discussion of this method, see \citet{Jorgensen02}. DUSTY uses $Y$
(=$R_{\rm{outer}}/R_{\rm{inner}}$), $p$, the power law exponent of the
density gradient ($n \propto r^{-p}$), and $\tau_{100}$, the opacity
at 100 $\mu$m as its free parameters. The temperature at the inner boundary was taken to be 250 K.  For the SED, fluxes were used at
60 and 100 $\mu$m \citep[IRAS,][]{Henning93}, 850 $\mu$m \citep[SCUBA
archive,][]{diFrancesco08}, 870 $\mu$m (LABOCA, this work) and 1.3 mm
\citep[SEST][]{Henning93}. No emission from MIPS at 24 $\mu$m
or ISO-SWS \citep{Nisini02,NoriegaCrespo04} was used. The 24 $\mu$m flux is included in the figure as reference. Note that the model is unable to account for this high flux. Deviations from spherical symmetry (outflow cavities), additional shock emission \citep{Velusamy07} or larger inner holes \citep{Jorgensen05b} are often inferred to explain the observed high mid-IR fluxes that cannot be fitted with DUSTY. The radial
profile was determined in directions away from the southwest outflow
and cloud material, ignoring any emission in a 90$^\circ$ cone to the south-west. Results for the best-fitting envelope model to the 16 $L_\odot$ for the source luminosity can be
found in Table \ref{5:table:prop}. Temperature and density distributions are displayed in Figure \ref{5:dens}. The corresponding fits are shown in
the insets of Fig. \ref{5:fig:laboca}.

The envelope contains a large amount of cold gas ($\sim$5.1
M$_{\rm{\odot}}$) within the outer radius of 20,800 AU ($\sim$0.1 pc),
but with a significant fraction of the mass concentrated towards the
inner envelope due to the steep density profile ($p$=1.8). The H$_2$
density at 1,000 AU is 2.4$\times$10$^6$ cm$^{-3}$. The transition from envelope to parental cloud is not taken into account.

\placeTableChapterFiveThree

\subsection{Envelope - gas}
 The physical structure of the gas is best traced by optically
 thin emission that probes the quiescent envelope gas at high density
 \citep{Jorgensen05}. To that end, the temperature and density
 structure derived from the dust radiative transfer model was used as a
 model input for the model using the 
 RATRAN radiative transfer code \citep{Hogerheijde00} with data files
 from the LAMDA database \citep{Schoeier05}. Two different scenarios were investigated . In the first, only freeze-out is taken into account, with
 an inner CO abundance $X_0$ of 2.7$\times10^{-4}$ with respect to
 H$_2$ and an outer CO abundance $X_d$ of 10$^{-5}$. In the second
 , the abundances are the same, except that a high abundance
 $X_0$=2.7$\times10^{-4}$ has been adopted for the outermost envelope
 regions where density is lower than $n_{\rm{de}}$=10$^5$
 cm$^{-3}$. This so-called 'drop' abundance profile is motivated by
 the fact that at low densitites, the timescales for freeze-out onto
 the grains are longer than the typical lifetimes of the cores of a
 few 10$^5$ yr.  The abundances $X_0$ and $X_d$ were derived by
 \citet{Jorgensen02} and \citet{Jorgensen05} for a range of sources,
 based on emission from optically thin lines. Isotope ratios were
 taken from \citet{Wilson94} of 550 for CO:C$^{18}$O and 70 for
 CO:$^{13}$CO. The velocity field of the envelope is represented by a
 turbulent width of 0.8 km s$^{-1}$.
 The gas and dust temperature were assumed to couple throughout the envelope. Gas and dust temperatures decouple in the cold outer envelope region with the gas temperature dropping by a factor two with respect to the dust temperature \citep{Ceccarelli96,Doty97}. This drop can be counteracted, however, by UV radiation from the outside impinging on the envelope and heating the gas. For a more thorough discussion, see also \citet{Jorgensen02}. Any small temperature difference in the outer envelope affecst mostly the low-$J$$^{12}$CO lines.  

Table \ref{5:tab:coprop} gives the results for these models. For each
model, the envelope contribution of both $\int T_{\rm{MB}}$ and
$T_{\rm{peak}}$ are given. The optically thin lines (C$^{18}$O 3--2
and 6--5, as well as the $^{13}$CO 8--7) show that the inclusion of a
higher abundance outer layer, as in a drop model, is necessary to
increase the C$^{18}$O 3--2 emission to the observed levels. At the
same time this does not change the emission of the C$^{18}$O 6--5
line. The adopted abundances of the drop abundance (Model 2) agree
best with the observed intensities of the optically thin lines,
including the upper limits on C$^{18}$O 6--5 and $^{13}$CO 8--7 within
the uncertainties.

\placeTableChapterFiveFour

 The very optically thick low-$J$ $^{12}$CO (2--1, 3--2 and 4--3)
lines cannot be fitted by either model due to the lack of self-absorption
in the observed lines, which is strongly present in the modelled
profiles.  To obtain a rough correction for self-absorption, gaussian
fits were made to the line wings of CO, similar to
the one in van Kempen et al. submitted.. Such gaussian fits provide only upper
limits to the peak emission of CO, as the true CO emission is best fitted with an infall velocity \citep{Schoeier02}.

Even with gaussian fits, the high-$J$ (6--5 and higher) lines of both
$^{13}$CO and $^{12}$CO are severly underproduced by almost a factor
3. Note that their model
line profiles do not show self-absorption. The emission in these lines
is almost identical in the drop or jump abundance models. Increasing
or decreasing the CO abundances $X_0$ and $X_d$ is not possible since
the emission of optically thin lines such as C$^{18}$O 6--5 and 3--2
would then either be over- or underestimated.
The main conclusion from the envelope models is therefore that an
additional, relatively optically thin but hot component unobscured by
the warm envelope region is needed to account for this emission. The
origin of this hot component producing narrow highly excited CO line
emission will be discussed in $\S$ 5.6. 

The integrated [C I] 2--1 intensity (2.3 K km s$^{-1}$) can be
reproduced with a constant abundance ratio C/H$_2$ of
3--5$\times$10$^{-7}$ (or about 0.1-3\% of CO, depending on radius),
typical for the densest molecular clouds.  The [C I] line is optically
thin, even for much higher abundances.  A C abundance as high as
5$\times$$10^{-6}$ can be maintained by photodissociation of CO due to
cosmic-ray induced UV photons deep inside the envelope
\citep[e.g.,][]{Flower94}.

HCO$^+$ 4--3 shows little to no difference between Model 1 and 2,
because most of the emission traces gas denser than 10$^5$
cm$^{-3}$. There is a significant difference in $T_{\rm{peak}}$ from
the line profile vs.\ gaussian fits for main isotope line because of
self-absorption.  The optically thin H$^{13}$CO$^+$ is slightly
underproduced, similar to the C$^{18}$O 3--2 peak temperature.

\subsection{Surrounding cloud material}

The envelope of HH~46 is surrounded by cold, quiescent cloud material
as evidenced by the LABOCA map and by the CO lines to the south-east
(e.g., 20$''$, -40$''$).
If the cloud is assumed to be isothermal and homogeneous, the CO
emission maps from 2--1 to 7--6 can be used to constrain the its
properties with the radiative transfer program RADEX
\citep{vanderTak07}\footnote{RADEX is available online at
http://www.sron.rug.nl/~vdtak/radex/radex.php}.  RADEX calculates the non-LTE excitation and line emission of molecules for a given temperature and density using an escape probability formulation for the radiative transfer. For gas at a constant temperature and density, such as in the surrounding cloud, the results from RADEX and RATRAN are comparable, but RADEX is easier and faster to use. A line width
of 1.2 km s$^{-1}$ is used from the gaussian fits to the CO 3--2
emission at the cloud positions. The ratios of CO 3--2/6--5 peak
temperatures at these cold cloud positions are at least 20, since CO
6--5 is not detected. RADEX simulations show that the CO 6--5 line has
to be sub-thermally excited, which can be done when the gas is at
low densities or at very low ($T<$10 K) kinetic temperatures. The line ratios are best fitted by a cloud with a
temperature of $\sim$14 K, a low density of a few times 10$^{3}$
cm$^{-3}$ and a CO column density of 10$^{17}$-10$^{18}$
cm$^{-2}$. 
The spatial
distribution of the CO 3--2 and 6--5 lines also indicate that these
cloud conditions extend homogeneously to at least 40$''$ south of HH
46, the extent of both CO 3--2 and 6--5 maps. Maps of the optically
thin C$^{18}$O 3--2, as have been done by \citet{vanKempen08} for
other sources, are needed to further constrain the column density and
spatial structure of the surrounding cloud.

\placeTableChapterFiveFive

\placeFigureChapterFiveSeven

\section{Outflow}

\subsection{Outflow temperature}

The spectral line maps clearly reveal the red-shifted outflow to the
south-west and the blue-shifted outflow to the north-east. The large
difference in extent between the two outflow lobes, already noted by
\citet{Olberg92}, is seen in all transitions, with the blue-shifted
lobe producing much weaker emission. The brighter and larger outflow
lobe seen with {\it Spitzer} \citep{NoriegaCrespo04,Velusamy07}
corresponds to the red-shifted outflow, which dominates the line
profiles of CO.  The lack of HCO$^+$ outflow wings as well as the
absence of HCO$^+$ emission in most positions except around the central envelope, suggest that the
swept-up gas is at a density of a few times 10$^{4}$ cm$^{-3}$ or
lower. The presence of a strong quiescent component at positions
associated with the outflow is discussed in $\S$ 6.

The wings of the isotopic lines provide an upper limit to the
optical depth, $\tau_{\rm{wing}}$, of the outflow. The ratio of
$T_{\rm MB}$ of the $^{12}$CO 3--2/$^{13}$CO 3--2 line wings at the
source position is shown in Fig. \ref{5:fig:CO_optdepth} as a function
of velocity. For a constant density of 3$\times$10$^{4}$ cm$^{-3}$ the
observed ratios correspond to $\tau_{\rm{wing}}$ of 1.8 (ratio=10) and
1 (ratio=25).  Fig. \ref{5:fig:CO_optdepth} shows that the line ratio
increases for more extreme velocities, introducing a dependency of
$\tau_{\rm{wing}}$ on velocity. In addition, there is a large jump in
$\tau_{\rm{wing}}$ from 6 to 7 km s$^{-1}$, representing the
transition between the optically thick quiescent and more optically
thin shocked material.  In the following analysis, it is 
assumed that all shocked outflow emission ($>$7 km s$^{-1}$, 1.7 km s$^{-1}$ with respect to the systemic velocity of 5.3 km s$^{-1}$) is
optically thin for all transitions. The effects of the moderate
optical depth of the lines are subsequently discussed.

\placeFigureChapterFiveEight
\placeFigureRebin
\placeFigureChapterFiveNine

Fig. \ref{5:fig:co_ratio} shows the ratios of the CO 3--2/6--5 main beam
antenna temperatures of line wings as functions of velocity for four
different positions along the red outflow axis ((0,0), (-20,-20),
(-30,-30) and (-40$''$,-35$''$)). The CO 6--5 data have not been binned to the larger CO 3--2 beam, so the comparison assumes similar volume filling factors of the shocked gas.  Figure \ref{5:rebin} shows the CO 6--5 spectra binned with a 10$''$ beam and with a 20$''$ beam. As can be seen, the differences between the spectra is negligible. Additional testing at other positions confirmed that the spectra do not differ by more than 20$\%$. Ratios are only plotted if the
emission in both wings is larger than 3$\sigma$.  

The kinetic temperature $T_{\rm{kin}}$ of the outflow can be derived
by comparing the intensity ratios in the line wings from various
transitions with model line intensities of \citet{vanderTak07}.  With
the density assumed to be constant at a few times $10^4$ cm$^{-3}$,
the observed 3--2/6--5 ratios of 2--3 correspond to kinetic
temperatures of about 120 to 150 K. Outflow emission is slightly
subthermally excited, especially the CO 6--5, with $T_{\rm{ex}}$ ranging from 85 to 120 K.  The rising ratios observed at the more extreme
velocities in Fig. \ref{5:fig:co_ratio} correspond to lower kinetic
temperatures, but even the highest ratios of $\sim$7 still indicate
kinetic temperatures greater than 70 K.  The variation of the optical depth
with velocity, as seen in Fig. \ref{5:fig:CO_optdepth}, could account
for the rising ratios seen in Fig. \ref{5:fig:co_ratio}, since a higher
optical depth will result in a lower ratio for the same
temperature. Even with the limit of $\tau_{\rm{wing}}$=1.7, the inferred kinetic
temperatures will drop by only 20$\%$ (see Fig.\ 4 of
\cite{Jansen96}).

Fig. \ref{5:fig:co_temp} presents the kinetic temperatures as functions
of position along the red outflow for a constant density of 2$\times$10$^4$ cm$^{-3}$. There is a clear trend towards
lower temperatures at larger radii.  Averaged between 7 and 9 km
s$^{-1}$, temperatures drop from 170 K close to the source to 80 K at
a distance 40$''$. However, the error bars, derived from the minimum
and maximum ratio at velocities greater than 7 km s$^{-1}$ show that
large variations are possible. In addition to the optical depth effects
discussed above,
it is also possible that the higher ratios at more extreme velocities
and larger distances correspond to lower densities $n$(H$_2$). The low
density of a few times 10$^3$ cm$^{-3}$ inferred for the surrounding
cloud raises the inferred kinetic temperatures from 80~K to 100--120
K. Such a drop in density significantly would lower the $T_{\rm{ex}}$. The current observational data cannot distinguish between these
two possibilities, but both options are consistent with a high kinetic outflow
temperature of $\sim$100 K that is constant with distance from the
source. 
This temperature is significantly higher than the $T_{\rm{kin}}$ and $T_{\rm{ex}}$
of $\sim$15 K assumed by \citet{Olberg92} for the CO 2--1/1--0
intensity ratios for the HH 46 outflow. Although some studies find $T_{\rm{ex}}$ that low \citep{Bachiller01}, recent studies, at times including high-$J$ CO, also find higher $T_{\rm{ex}}$ and $T_{\rm{kin}}$ \citep[e.g.][]{Hirano01, Lee02}. 
While a cooler outflow component is not excluded, our data
clearly show the presence of warmer outflow gas. Kinetic temperatures
as low as 50 K would require outflow densities in excess of $10^5$
cm$^{-3}$, which are excluded by the HCO$^+$ data.

In a similar analysis, the blue-shifted outflow has kinetic temperatures of 70
to 100 K with a lower limit of 50 K, somewhat cooler than the red
lobe. In addition to a variation with velocity, the ratios also seem
to vary with distance from the source. At larger distances, the CO
3--2/6--5 intensity ratio is almost a factor of 2 higher than at the
source position.

\citet{Hatchell99} use a swept-up shell model to predict kinetic
temperatures along the outflow axis and walls. Their predicted values
of 50 to 100 K agree very well with our derived temperatures in both
outflow lobes (Fig. \ref{5:fig:co_temp}). Such temperatures are much
lower than calculated for entrainment models, which predict $>$ 1000 K
\citep[e.g.,][]{Lizano95}. \citet{Arce02} compare different outflow
models and show that at least some flows are best explained with the
jet-driven bow shock model. For a thorough review of outflow models, see \citet{Arce07}. The models
from \citet{Hatchell99} predict an almost  constant temperature along most
of the outflow axis with increasing temperatures near the bow shock,
the main site of energy deposition. This is consistent with our
constant temperature along the outflow axis.

\subsection{Other outflow properties}

Additional properties of both outflow lobes are derived from the
molecular emission maps following the recipe outlined in
\citet{Hogerheijde98}, in which the radii, masses, dynamical time
scales, outflow force and kinetic luminosity are calculated (see Table
\ref{5:table:out}). The results are corrected for the inclination of
35$^\circ$ found for this source \citep{Reipurth91,Micono98} using an
average of the three correction factors from \citet{Cabrit90} (their
Figures 5--7). These factors range from 1.2 to 2.5, especially for the
mass, and are introduced to account for the difference in observed
$V_{\rm{max}}$ on the sky and the actual extreme velocities. 
  Although excitation temperature variations are seen throughout the
  outflow that depend on velocity and distance from the source, we assume an
  average excitation temperature of 100 K for the red outflow lobe and 70 K for the blue outflow lobe
  for the derivation of these parameters.

The resulting values can be uncertain up to an order of magnitude due
to the variations in the covered area and thus in radius, especially
for the CO 4--3 (observed with FLASH).  The mass estimates of the low
excitation CO lines (e.g., the CO 2--1) may be overestimated by up to
a factor 4, due to the larger area covered and the assumption that the
outflow is iso-thermal at 100 K. CO 2--1 may be dominated by 
cooler gas with temperatures down to, say, 50 K, lowering the mass
estimate to a lower limit 0.8 M$_{\rm{\odot}}$, only 25$\%$ higher
than the masses found for the high-$J$ CO lines. Temperature
differences are likely also responsible for the (smaller) difference in
masses in the other lines. 

Even with the uncertainties in covered area, the dynamical time scales
for these outflows all converge on 10,000 to 20,000 years, with no
difference between the red and blue outflow lobes. The values of Table
\ref{5:table:out} are similar to the results found in \citet{Cabrit92}
and \citet{Hogerheijde98} for other Class I outflows and also agree
with the results from \citet{Olberg92} who find
$L_{\rm{kin}}=$4.5$\times$10$^{-3}$ L$_\odot$ and dynamical time
scales of 4$\times$10$^{4}$ yr. Although our dynamical time scales are
a factor of 2 smaller, this can be accounted for by the smaller
covered area in our observations.

 The main discrepancy with older studies is that our outflow mass is
 up to an order of magnitude higher than that by \citet{Olberg92}. The
 origin of this difference is two-fold. First, \citet{Olberg92} do not
 apply correction factors from \citet{Cabrit90}. Second, there is a
 large difference between assumed temperatures in the red lobe (15 K
 vs 100 K). Although \citet{Olberg92} derive their temperature from
 the CO 1--0 and 2--1 emission, the presence and intensity of CO 4--3,
 6--5 and 7--6 outflow emission strongly constrain temperatures to our
 higher value of 100 K.
As illustrated by the CO 2-1 example, this can introduce a significant
difference in the masses.

 At first sight, there is no difference with the masses derived by
 \citet{Chernin91}, but they assumed a similar temperature of 15-30 K
 as \citet{Olberg92}.  However, \citet{Chernin91} took an optical
 depth of $\sim$5 for the outflow emission, which is not confirmed by
 our observed lack of outflow emission in the $^{13}$CO lines.  For
 our observed maximum $\tau_{\rm wing}$=1.7, the outflow mass as
 derived from CO 3--2 would increase by a factor of 2 ($\approx \tau /
 (1-e^{-\tau})$). 

 From the dust map, a total (envelope + cloud) mass of 8
  M$_\odot$ is derived (see $\S$ 5.3.2). The total outflow mass in Table
  \ref{5:table:out} can be as high as 3.2 M$_\odot$ and thus consist of a significant portion
  (40$\%$) of that total mass.  If the low mass for CO 2--1 is
  adopted, this percentage drops to 10$\%$.

\citet{Bontemps96} empirically derive a relation between the flow
force, $F_{\rm{CO}}$ and bolometric luminosity
\begin{equation}
\log(F_{\rm{CO}})=0.9\log(L_{\rm{bol}})-5.6
\end{equation} 
and the envelope mass 
\begin{equation}
\log(F_{\rm{CO}})=1.1\log(M_{\rm{env}})-4.15
\end{equation}

 Using the bolometric luminosity of 16 $L_\odot$ and envelope mass of
 $\sim$5 M$_\odot$, flow forces of between 3$\times$10$^{-5}$ and
 4$\times$10$^{-4}$ M$_\odot$ yr$^{-1}$ km s$^{-1}$ are expected.  The
 parameters for the red outflow lobe indeed show flow forces for most
 CO transitions that agree with these predictions. Only the flow force
 of CO 2--1 is a factor of 2.5 higher. If the lower mass of 0.8 M$_\odot$
 is used for this transition due to a lower outflow temperature of 50
 K, the flow force similarly drops by a factor of 3.5, almost identical
 to the flow forces found for the CO 6--5 and 7--6.

The observed flow forces for the blue
outflow lobe are up to two orders of magnitude lower than those of the
red lobe. This $`$missing' outflow force can be explained by assuming
that the blue-shifted outflow escapes the envelope and surrounding
cloud and thus interacts much less with the surrounding cloud due to a
lack of cloud material in the path of this outflow lobe.

\section{Origin of the quiescent high-$J$ CO line emission}
\subsection{Photon heating of cavity walls}

The analysis in $\S$ 4.2 shows that an envelope model derived from the
dust emission is not able to fit the quiescent emission of the
high-excitation lines by a factor of $\sim$2.5, even though the
low-excitation optically thin lines are well-fitted with standard
abundances.
The limits on the C$^{18}$O 6--5, combined with the information that
the $^{12}$CO 6--5 emission from the envelope is optically thick,
implies that an additional heating component must be present that
produces quiescent emission. There are several constraints on this
component.  First, it cannot originate in the inner regions of the
protostellar envelope or be obscured by the envelope itself, because
emission from such a component cannot escape through the optically
thick outer envelope emission.  Second, it must be (nearly) optically
thin, since the observed C$^{18}$O 6--5 limit is already reached by
the envelope itself.  Thus, it cannot contribute significantly in
mass.  Third, it must be extended since a similar warm quiescent
component is also clearly seen at other positions covering the
red-shifted outflow through narrow $^{12}$CO 6--5 emission, e.g.  at
positions (-20$''$10$''$) and (-20$''$, -20$''$)
(see Fig. \ref{5:fig:mapsspec} and \ref{5:total}). Observations of $^{13}$CO 6--5 at other positions also confirm that this quiescent emission cannot be very
optically thick ($\tau<0.4$).

\citet{Spaans95} investigated the influence of photon heating on the
emission of high-$J$ CO lines, in order to explain the bright but
narrow $^{12}$CO and $^{13}$CO 6--5 single spectra observations of
Class I sources \citep{Hogerheijde98}. In this process ultraviolet
(UV) photons heat the gas of the outflow cavity walls to temperatures
of a few hundred K, but are unable to dissociate the CO molecules. 
The photons in this model originate in a $\sim$10,000 K radiation
field of the boundary layer in the accretion disk. It was proposed that dust present in the cavity scatters the UV photons towards the cavity walls.
The turbulent velocity of the gas at the cavity walls is low, thus
explaining the narrow width of the CO 6--5 and 7--6 lines. This extra
warm gas due to photon heating could account for the differences in
the observed $^{12}$CO and $^{13}$CO 6--5 emission on source and that
modelled by just a protostellar envelope.

\placeFigureChapterFiveTen

Although \citet{Spaans95} only include photons created in the
accretion disk and thus limit their Photon Dominated Region (PDR) to a
central region of up to 4,000 AU in size, this model can be extended
if additional sources of UV photons are included. Such photons can be
produced at bow shock positions of the outflow if $J$-shocks are
present there \citep{Neufeld89}.  
 Such photons can also heat the dust at the cavity walls, increasing the emission of the dust particles at continuum wavelengths as presented in Fig. 2. The elongated continuum emission of HH46 on a 1$'$ scale in the south-east direction may indeed be caused by warm dust, instead of more massive cold dust or CO emission. 
\citet{Velusamy07} show that apart
from the 24 $\mu$m hotspot of HH 47C, two additional 24 $\mu$m
emission spots can be found within the outflow cavity of HH 46/47,
that likely originate in the jet \citep[see Fig. 3
in][]{Velusamy07}. These positions produce additional UV photons,
creating a much more extended PDR. The bow shock at the position of HH
47C, the two new jet/shock positions and the accretion shock boundary
layer combined are excellent sources of the necessary UV photons.
Figure \ref{5:cartoon} shows a schematic overview of the red-shifted
outflow cavity and the proposed process of photon heating within the
HH 46/47 outflow. $\S$ 6.2 discusses the path of the UV photons in more detail, while $\S$ 6.3 presents the constraints on the UV field.

\subsection{Importance of the outflow geometry}

To illuminate a larger area of the cavity wall than possible by direct irradiation \citet{Spaans95} invoked scattering of the UV photons by the dust present in the outflow cavity, even though such scattering is dominated by anisotropic, mainly forward scattering. With a constant opening angle (see inset {\bf a} of Fig. \ref{5:cartoon}), only a very small area of the envelope wall will be directly impacted by UV photons. However, with densities of a few 100 cm$^{-3}$ the mean free path of a UV photon until its first scattering event is $\sim$65,000 AU, with the assumption that on average an $A_v$ of 1 produces a single scattering event \citep{Draine03}.  This is significantly larger ($\sim$ 150$''$ on the sky) than the area covered by the outflow cavity. 

A parabolic shape of the outflow cavity in the regions near the envelope, as seen in inset {\bf b} of Fig. \ref{5:cartoon}, allows much more UV radiation from the accretion boundary layer to illuminate the cavity walls at  larger distances of a few thousand AU.  A more quantitative description of outflow shapes and its effect on the illumination of the outflow walls is beyond the scope of this paper.

\subsection{Constraining the UV field}
The temperature of the cavity walls can be derived from the radiation
field originating from these shocks. The radiation field $G_0$ at the
cavity walls can be characterized by
\begin{equation}
G_0 = \sum G_0^i f^i
\end{equation} 
with $G_0^i$ the radiation field originating at each shock position
$i$ and $f^i$ a geometric dilution factor to account for the
difference in the UV emitting surface to the total illuminated surface
of the cavity walls. A dissociative shock $i$ produces 3$\times$10$^7$
($v_s/100$ km s$^{-1}$)$^3(n_0)$ photons cm$^{-2}$ s$^{-1}$ assuming
all energy is emitted at 10 eV \citep{Neufeld89}. This is equivalent
of a $G_0^i$ of 30,000 assuming a density $n_0$ of 10$^{4}$ cm$^{-3}$
and a shock velocity $v_s$ of 220 km s$^{-1}$, derived for the bow
shock \citep{Fernandes00}.  The average factor of $f^i$ over all
shocks is assumed to be 0.005, meaning that the working surface of
each shock is $\sim$1/200 in size compared to the cavity surface.
However, not all shocks will have a velocity as high as 220 km
s$^{-1}$. It is more likely that the velocities of the secondary
shocks within the cavity are much lower. But similarly, the size of
the working surfaces of the shocks, and thus the $f^i$ can be larger,
resulting in a comparable overall strength $G_0$, but of a different
`color'. With four spots, the cavity walls are illuminated by a total
$G_0$ equaling $\sim$600 with an uncertainty of a factor of 2 to 3 due
to the uncertainty in the $f^i$. Using the results from \citet[ their
  Figure 1]{Kaufman99}, the PDR surface temperature is then
constrained to 250--400 K, sufficient to produce the observed lines.

[C I] 2--1 is detected near and at the bow shock position, but not at the outflow positions closer to the star, although the dynamic range in the [C I] data is small. This could indicate
that the penetrating UV photons within the region of the outflow lobe closer to the star are not able to dissociate CO
significantly, constraining the color temperature of the radiation
field. \citet{Neufeld89} show that shocks with velocities less than 90
km s$^{-1}$ do not produce CO dissociating photons.  If all shocks
have $v_s \approx$80 km s$^{-1}$ the estimated $G_0^i$ drops by a
factor $\sim$20.  In that case, the total $G_0$ is not sufficient to
heat the cavity walls to surface temperatures of 250--400 K, but only
to about 100 K.  Lower velocities are likely for the shocks observed
inside the cavity. However, the known shock velocity of 220 km s$^{-1}$ \citep{Fernandes00} is sufficient to produce CO dissociating photons, consistent with the observed narrow [C I] emission at this position. This situation is
reminiscent of the observation of strong quiescent [C I] 1-0 emission
in the supernova remnant IC 443 ahead of the shock, originating from
 photodissociation of CO in the pre-shocked gas \citep{Keene96}. 
The likely scenario for HH~46 is thus
\begin{enumerate}
\item Non-dissociative UV photons are created in the boundary layer and secondary shocks, while the bow shock produces mainly CO dissociative UV photons
\item Along the outflow axis, the cavity walls are heated to sufficient temperatures to produce the quiescent high-$J$ CO emission.
\item Closer to the bow shock, dissociation of CO becomes significant in addition to the heating of the cavity walls, explaining the lack of CO 6--5 emission and presence of [C I] 2--1 emission near and at the bow shock
\end{enumerate}

Observations of far-IR CO lines, such as are possible with the $HIFI$
and {\it PACS} instruments on {\it Herschel}, are needed to constrain
the exact temperature of CO gas at the cavity walls.  

Slow ($v_s$ = 5-10 km s$^{-1}$) $C$-type shocks along the outflow
cavity walls are able to generate similar amounts of CO emission
\citep{Draine84}. However, the narrow nature of the line profile, as
well as the presence of CO 6--5 emission over the entire area traced
by the IR outflow, make this scenario less likely than the photon
heating, although it could contribute some.

\section{Conclusions}
In this paper, we characterize the structure of protostellar envelope
and the molecular outflow associated with HH~46 IRS~1, as well as its
immediate surrounding cloud material, through dust and molecular line
maps. Broad and narrow CO lines are observed ranging in transitions from 2--1 to 7--6,
including isotopologues. The three distinct components can be best
described by the following model: \\

\begin{itemize}
\item {\bf Envelope} - The envelope of HH~46 with $\sim$3--5 M$_\odot$ (within $T>10$ K) is
  one of the most massive ones found for a Class I low-mass protostar, but is
  densely concentrated toward the center ($n\propto r^{-1.8}$). The
  C$^{18}$O line emission from the envelope can be best fitted with a
  drop abundance of 2.7$\times$10$^{-4}$/1$\times$10$^{-5}$
  above/below 30 K and below/above 10$^{5}$ cm$^{-3}$. However, such
  abundanes are unable to reproduce the observed $^{12}$CO and
  $^{13}$CO 6--5 and 7--6 emission. The dense envelope itself is best
  traced by the HCO$^+$ 4--3 emission, which has very little outflow
  contribution and shows a spherical distribution. Densities in the
  inner few hundred AU of the envelope are high ($>$10$^7$ cm$^{-3}$),
  with high optical depths of the HCO$^+$ 4--3 and all $^{12}$CO
  lines. A C/H$_2$ abundance of a few times 10$^{-7}$ is found, which
  can be maintained by photodissociation of CO by
cosmic ray induced UV photons.\\

\item {\bf Surrounding cloud} - The surrounding cloud  extends over more than 100$''$ to the south-southwest but does not extend
further than $\sim$30$''$ north of HH 46 IRS 1 where even no low
excitation CO emission is found. Cloud conditions include a low
density of a few times 10$^3$ cm$^{-3}$, derived from limits on the CO
6--5 emission at positions such as (30$''$,-20$''$). The total column
of CO is $\sim$10$^{18}$ cm$^{-2}$.\\

\item {\bf Outflow} - The red-shifted molecular outflow, extending at
  least 40$''$ from the source, produces strong molecular line wings
  up to CO $J$=7--6 and heats the surrounding cloud and envelope
  significantly close to the star.  Spatially, the red-shifted outflow
  lobe corresponds to the bright infrared outflow lobe from
  \citet{NoriegaCrespo04}.  Optical depth of the CO 3--2 outflow wing
  is less than 1.7. Kinetic temperatures of the red-shifted outflow
  are of order 100--150 K close to the star for flow densities of
  2$\times10^4$ cm$^{-3}$, but drop to 80 K further from the central
  source if densities and optical depth remain constant.  However, the
  data are also consistent with a constant kinetic temperature in the
  covered area if densities decrease to a few 10$^3$ at a distance
  $>$40$''$ from the central source, as found for the surrounding
  cloud.

\quad  Temperatures of both outflow lobes are significantly higher than
the previously derived temperature of 15 K, but agree well with the
model predictions of \citet{Hatchell99} for a swept-up shell
model. The high temperature causes the observed outflow mass to be
significantly higher (almost an order of magnitude) than derived in
older studies such as \citet{Olberg92}.  Bright narrow [C I] is found near the bow shock, indicating that the bow shock produces CO dissociating photons.\\

\item {\bf Origin of high-$J$ CO} - The emission seen in the higher
excitation CO transition has three main origins.
\begin{enumerate} 
\item The dense envelope produces optically thick emission in both CO
  6--5 and 7--6, originating in the warm ($T > $ 50 K) inner envelope,
  accounting for roughly 1/3 of the observed line intensities on
  source.
\item High-$J$ CO emission is detected in the red- and blue-shifted
  outflow wings at some positions along the outflow axis.
\item The bulk of the high-$J$ CO emission has narrow lines
and is produced by photon heating.  UV photons originating in the bow
shocks, jet shocks and accretion boundary layer
heat the cavity walls up to a few hundred K. 
The lack of strong associated [C I] emission near the source indicates that the UV
photons do not photodissociate CO, suggesting shock velocities lower
than 90 km s$^{-1}$ such as could be present inside the
cavity. CO dissociating photons are limited to the region close to the bow shock. 
\end{enumerate}
\end{itemize}
This paper shows that the addition of the high-$J$ CO emission lines
as observed with CHAMP$^+$ provides new insights into the physical
structure of the protostellar envelope and molecular outflow of
HH~46. The presence of narrow line emission in the high-$J$ CO lines
throughout the outflow suggests that photon heating is an important
process in HH~46. The high-$J$ isotopic CO lines, in particular those
of C$^{18}$O 6--5, have been key in constraining the envelope
model.The UV radiation
implied by the high-J CO and [C I] observations should also have
 significant consequences for the chemistry of other species and should
 enhance radicals like CN and OH along the outflow walls. Future high-frequency observations using high=$J$ CHAMP$^+$, Herschel and, in the
long run, ALMA high frequency bands, will provide unique constraints on the
interaction between outflow and envelope.

\begin{acknowledgements}

 TvK and astrochemistry at Leiden Observatory are supported by a
 Spinoza prize and by NWO grant 614.041.004. CHAMP$^+$ is built with NWO grant 600.063.310.10. TvK is grateful to the
 APEX staff for carrying out the bulk of the low-frequency
 observations. Carlos de Breuck is thanked for providing the APEX-1
 observations on a very short notice within the science verification
 project E-81.F-9837A. We appreciate the input of Steve Doty into an illuminating and essential discussion about outflow cavity shapes. We are grateful for support from Ronald Stark throughout construction of CHAMP$^+$. Constructive comments by an anonymous referee helped improve the paper.

\end{acknowledgements}
\bibliographystyle{../../bibtex/aa}
\bibliography{../../biblio}

\end{document}